\documentclass[11pt,a4paper]{article}

\usepackage[bookmarks=true,bookmarksnumbered=true,setpagesize=false]{hyperref}

\usepackage{ascmac}

\usepackage[utf8]{inputenc}
\usepackage{amsmath,amsfonts,amssymb,amsthm,color,enumerate}
\usepackage{graphicx}
\usepackage{slashed}
\usepackage{color}
\usepackage[dvipsnames]{xcolor}

\usepackage{comment}
\usepackage{a4wide}
\usepackage{cancel} 
\usepackage[normalem]{ulem} 

\newcommand{\tx}{\text}
\newcommand{\ov}{\over}

\newcommand{\R}{\tx{R}} 

\newcommand{\paren}[1]{\left(#1\right)}
\newcommand{\fn}[1]{\!\left(#1\right)}
\newcommand{\sqbr}[1]{\left[#1\right]}

\newcommand{\ab}[1]{\left|#1\right|}
\newcommand{\df}{\text{d}}
\newcommand{\ol}{\overline}
\newcommand{\al}[1]{\begin{align}#1\end{align}}
\newcommand{\als}[1]{\begin{align*}#1\end{align*}}

\newcommand{\h}{\hat}

\newcommand{\mc}[1]{\ensuremath{\mathcal{#1}}}
\newcommand{\bs}[1]{\ensuremath{\boldsymbol{#1}}}

\DeclareMathOperator{\BR}{BR}

\newcommand{\GeV}{\,\text{GeV}}
\newcommand{\TeV}{\,\text{TeV}}
\newcommand{\fb}{\,\text{fb}}
\newcommand{\pb}{\,\text{pb}}

\renewcommand{\L}{\text{L}}

\begin{document}

\title{
\vbox{
\baselineskip 14pt
\hfill \hbox{\normalsize KIAS-P17006}\\
\hfill \hbox{\normalsize OU-HET/923}\\
}		
\vfill
{\bf Di-higgs enhancement by neutral scalar\\
as probe of new colored sector}
}

\author{
Koji Nakamura,\thanks{\tt koji.nakamura@cern.ch} $^1$\
Kenji Nishiwaki,\thanks{\tt nishiken@kias.re.kr} $^2$\
Kin-ya Oda,\thanks{\tt odakin@phys.sci.osaka-u.ac.jp} $^3$\\
Seong Chan Park,\thanks{\tt sc.park@yonsei.ac.kr} $^{2,4}$ and
Yasuhiro Yamamoto\thanks{\tt yamayasu@yonsei.ac.kr} $^4$\bigskip\\
$^1$\it\normalsize
IPNS, High Energy Accelerator Research Organization (KEK), Ibaraki 305-0801, Japan\\
$^2$\it\normalsize
School of Physics, Korea Institute for Advanced Study (KIAS), Seoul 02455, Republic of Korea\\
$^3$\it\normalsize
Department of Physics, Osaka University, Osaka 560-0043, Japan\\
$^4$\it\normalsize
Department of Physics \& IPAP, Yonsei University, Seoul 03722, Korea\\
	}
\date{\today}
\maketitle

\begin{abstract}\noindent
We study a class of models in which the Higgs pair production is enhanced at hadron colliders by an extra neutral scalar. 
The scalar particle is produced by the gluon fusion via a loop of new colored particles, and decays into di-Higgs through its mixing with the Standard Model Higgs.
Such a colored particle can be the top/bottom partner, such as in the dilaton model, or a colored scalar which can be triplet, sextet, octet, etc., called leptoquark, diquark, coloron, etc., respectively. We examine the experimental constraints from the latest Large Hadron Collider (LHC) data, and discuss the future prospects of the LHC and the Future Circular Collider up to 100\,TeV.
We also point out that the $2.4\,\sigma$ excess in the $b \bar{b} \gamma \gamma$ final state reported by the ATLAS experiment can be interpreted as the resonance of the neutral scalar at 300\,GeV. 
\end{abstract}


\newpage

\section{Introduction}

The di-Higgs production will continue to be one of the most important physics targets in the Large Hadron Collider (LHC) and beyond, since its observation  leads to a measurement of the tri-Higgs coupling, and will provide a test if it matches with the Standard Model (SM) prediction~\cite{Glover:1987nx,Eboli:1987dy,Dawson:1998py,Baur:2002qd,Baur:2003gp,Shao:2013bz,Grigo:2013rya,deFlorian:2013jea,Degrassi:2016vss,Borowka:2016ypz,Ferrera:2016prr}.
Since its production in the SM is destructively interfered by the top-quark box-diagram contribution, sizable production of di-Higgs directly implies a new physics signature~\cite{Baglio:2012np}.

It is important to examine in what kind of a model the di-Higgs signal is enhanced.
Indeed the enhancement has been pointed out in the models with two Higgs doublets~\cite{Craig:2013hca,Baglio:2014nea,Aad:2014yja,Hespel:2014sla,Barger:2014qva,Lu:2015qqa,Dorsch:2016tab,Kling:2016opi,Bian:2016awe}, type-I\hspace{-.1em}I seesaw~\cite{Han:2015sca},
light colored scalars~\cite{Kribs:2012kz},
heavy quarks~\cite{Dawson:2012mk},
effective operators~\cite{Pierce:2006dh,Kanemura:2008ub,Dolan:2012rv,Nishiwaki:2013cma,Chen:2014xra,Liu:2014rba,Slawinska:2014vpa,Goertz:2014qta,Azatov:2015oxa,Lu:2015jza,Carvalho:2015ttv,deFlorian:2016uhr,Gorbahn:2016uoy,Carvalho:2016rys,Cao:2016zob},
dilaton~\cite{Dolan:2012ac},
strongly interacting light Higgs 
and minimal composite Higgs
~\cite{Contino:2010mh,Grober:2010yv,Contino:2012xk,Grober:2016wmf},
little Higgs~\cite{Liu:2004pv,Dib:2005re,Wang:2007zx},
twin Higgs~\cite{Craig:2015pha},
Higgs portal interactions~\cite{Dolan:2012ac,Chen:2014ask,Robens:2015gla,Martin-Lozano:2015dja,Falkowski:2015iwa,Buttazzo:2015bka,Dawson:2015haa,Robens:2016xkb,Dupuis:2016fda,Banerjee:2016nzb},
supersymmetric partners~\cite{Plehn:1996wb,Djouadi:1999rca,Baur:2003gp,Cao:2013si,Han:2013sga,Bhattacherjee:2014bca,Cao:2014kya,Djouadi:2015jea,Batell:2015zla,Wu:2015nba,Batell:2015koa,Costa:2015llh,Agostini:2016vze,Hammad:2016trm,Biswas:2016ffy},
and Kaluza-Klein graviton~\cite{Khachatryan:2016sey}.
Other related issues are discussed in Refs.~\cite{Asakawa:2010xj,Papaefstathiou:2012qe,Klute:2013cx,Goertz:2013kp,Barr:2013tda,Chen:2013emb,Papaefstathiou:2015iba,vonBuddenbrock:2015ema,Cao:2015oaa,Cao:2015oxx,Huang:2015izx,Behr:2015oqq,Cao:2016udb,Kang:2016wqi,vonBuddenbrock:2016rmr,Fichet:2016xvs,Fichet:2016xpw,Huang:2017jws}.
The triple Higgs productions at the LHC and the future circular collider (FCC) are also discussed in Refs.~\cite{Plehn:2005nk,Maltoni:2014eza,Papaefstathiou:2015paa}.

In this paper, we study a class of models in which the di-Higgs process is enhanced by a resonant production of an extra neutral scalar particle. Its  production is radiatively induced by the gluon fusion via a loop of new colored particles. Its tree-level decay is due to the mixing with the SM Higgs boson. As concrete examples of the new colored particle that can decay into SM ones in order not to spoil cosmology, we examine the top/bottom partner, such as in the dilaton model, and the colored scalar which are triplet (leptoquark), sextet (diquark), and octet (coloron).

We are also motivated by the anomalous result reported by the ATLAS Collaboration: the 2.4\,$\sigma$ excess in the search of di-Higgs signal using $b\bar b$ and $\gamma\gamma$ final states with the $m_{(b\bar{b})(\gamma\gamma)} (=m_{hh})$ invariant mass at around 300\GeV~\cite{Aad:2014yja}.
The excess in $m_{(\gamma\gamma)}$ distribution is right at the SM Higgs mass on top of both the lower and higher mass-side-band background events.
The requested signal cross section roughly corresponds to 90 times larger than what is expected in the SM.
Thus the enhancement, if from new physics, should be dramatically generated via e.g.\ a new resonance at $300\GeV$.

This paper is organized as follows. In Sec.~\ref{model section}, we present the model.
In Sec.~\ref{signal section}, we show how the di-Higgs event is enhanced.
In Sec.~\ref{constraint}, we examine the constraints on the model from the latest results from the ongoing LHC experiment. 
In Sec.~\ref{2.4sigma}, we present a possible explanation for the 2.4\,$\sigma$ excess.
In Sec.~\ref{summary section}, we summarize our result and provide discussion.
In Appendix~\ref{general model}, we show how the effective interaction between the new scalar and Higgs is obtained from the original Lagrangian.
In Appendix~\ref{Z2 model section}, we give a parallel discussion for the $Z_2$ model.
In Appendix~\ref{colored scalar section}, we spell out the possible Yukawa interactions between the colored scalar and the SM fields.
%
\section{Model}\label{model section}

\begin{figure}[tp]
\begin{center}
\includegraphics[width=0.35\textwidth]{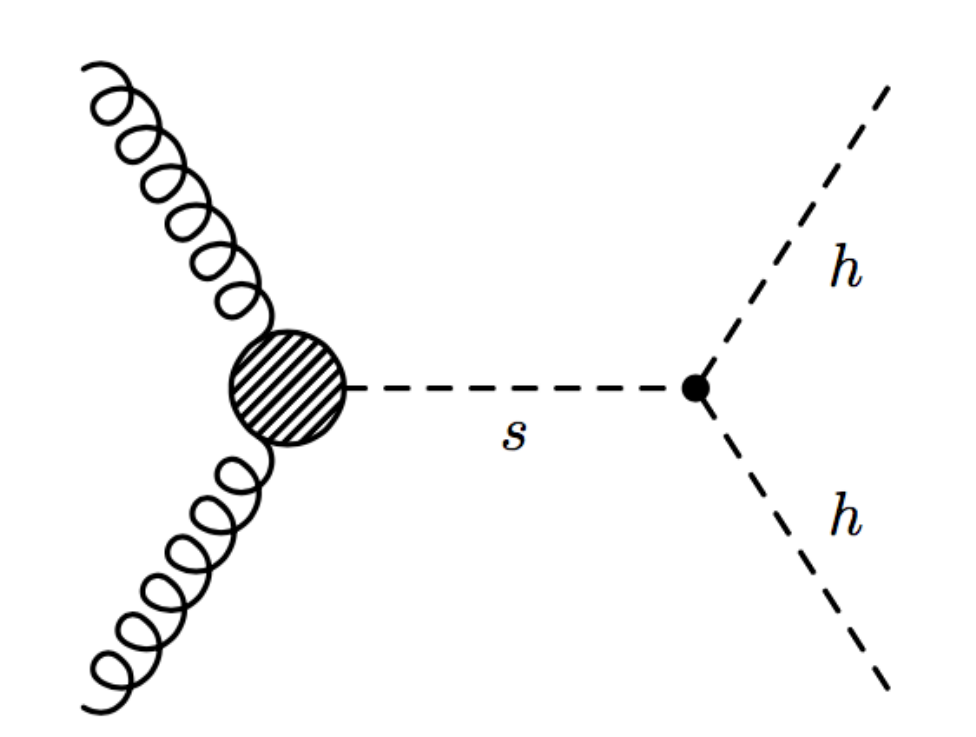}
\caption{Di-Higgs ($hh$) production mediated by $ s$.}
\label{Di-Higgs figure}
\end{center}
\end{figure}

We consider a class of models in which the di-Higgs ($hh$) production is enhanced by the schematic diagram depicted in Fig.~\ref{Di-Higgs figure}, where
$s$ denotes the new neutral scalar and
the blob generically represents an effective coupling of $s$ to the pair of gluons via the loop of the extra heavy colored particles.
We assume that $h$ and $s$ are lighter and heavier mass eigenstates obtained from the mixing of the neutral component of the $SU(2)_L$-doublet $H$ and a real singlet $S$ that couples to the extra colored particles:
\al{
H^0	&=	{v+h\cos\theta+ s\sin\theta\over\sqrt{2}},
		\label{H0 written}\\
S	&=	f-h\sin\theta+ s\cos\theta,
		\label{S written}
}
where $\theta$ is the mixing angle and $v$ and $f$ denote the vacuum expectation values (VEVs):
\al{
\left\langle H^0\right\rangle
	&=	{v\over\sqrt{2}},	&
\left\langle S\right\rangle
	&=	f,
		\label{VEVs}
}
with $v\simeq246\GeV$ and $m_h=125\GeV$.
We phenomenologically parametrize the effective $shh$ interaction as
\al{
\Delta\mc L
	&=	-{\mu_\tx{eff}\sin\theta\over2} s h^2,
		\label{shh coupling}
}
where $\mu_\tx{eff}$ is a real parameter of mass dimension unity, whose explicit form in terms of original Lagrangian parameters is given in Appendix~\ref{general model}.
We note that the parameter $\mu_\tx{eff}$ is a purely phenomenological interface between the experiment and the underlying theory in order to allow a simpler phenomenological expression for the tree-level branching ratios; see Sec.~\ref{tree decay section}.
We note also that the $\theta$-dependent $\mu_\tx{eff}\fn{\theta}$ goes to a $\theta$-independent constant in the small mixing limit $\theta^2\ll1$; see Appendix~\ref{general model} for detailed discussion. 
In Sec.~\ref{constraint}, it will indeed turn out that only the small, but non-zero, mixing region is allowed in order to be consistent with the signal-strength data of the 125\,GeV Higgs at the LHC.

The extra colored particle that runs in the loop, which has been generically represented by the blob in Fig.~\ref{Di-Higgs figure}, can be anything that couples to~$S$. 
It should be sufficiently heavy to evade the LHC direct search and decay into SM particles in order not to affect the cosmological evolution.
In this paper, we consider the following two possibilities: a Dirac fermion that mixes with either top or bottom quark and a scalar that decays via a new Yukawa interaction with the SM fermions. For simplicity, we assume that the new colored particles are singlet under the SU(2)$_L$ in both cases.

In Table~\ref{table of fields}, we list the colored particles of our consideration. The higher rank representations of $SU(3)_C$ for the colored scalars are terminated at $\bs 8$ in order not to have too higher dimensional Yukawa operators.\footnote{
The ultraviolet completion of the higher dimensional operator requires other new colored particles.
 We assume that their contributions are subdominant.
 E.g.\ they do not contribute to the effective $ggs$ vertex if they do not have a direct coupling to $S$.
}
The triplet $\phi_{\bs 3}$ is nothing but the leptoquark. It is worth noting that the leptoquark with $Y=-1/3$ may account for $R_{D^{(*)}}$, $R_K$, and $\paren{g-2}_\mu$ anomalies simultaneously~\cite{Bauer:2015knc}.

\begin{table}[tp]
\centering
\begin{tabular}{|c|ccc|cccc|c}
\hline
			&\multicolumn{3}{c|}{Dirac spinor}&\multicolumn{4}{c|}{complex scalar}\\
field		& $T$		& $B$	& \dots	& $\phi_{\bs 3}$			& $\phi_{\bs 6}$	& $\phi_{\bs 8}$	&\dots \\
\hline
$SU(3)_C$	& \bs 3		& \bs 3	& \dots		& \bs 3		& \bs 6			& \bs 8 &\dots\\
$Q$			& $2\ov3$	& $-{1\ov3}$ & \dots	& $-{1\ov3}$, $-{4\ov3}$ & ${1\ov3}$, $-{2\ov3}$, ${4\ov3}$ & 0, $-1$&\dots\\
$\Delta b_g$	& ${2\ov3}$ & ${2\ov3}$ & \dots & ${1\ov6}$ & ${5\ov6}$ & $1$ & \dots\\
$\Delta b_\gamma$
	& ${16\ov9}$ & ${4\ov9}$ & \dots &
	${1\ov9}$, $16\ov9$ & $2\ov9$, $8\ov9$, $32\ov9$& 0, ${8\ov3}$&\dots\\
$\eta$	&	 & $y_FN_F{v\ov M_F}$ & & \multicolumn{4}{c|}{$\kappa_\phi N_\phi{fv\ov M_\phi^2}$}\\
\hline
\end{tabular}
\caption{Colored particles that may run in the loop represented by the blob in Fig.~\ref{Di-Higgs figure}, and their possible parameters. We assume that they are $SU(2)_L$ singlets. The electromagnetic charge $Q$ is fixed to allow a mixing with either top or bottom quark for the Dirac spinor and a Yukawa coupling with a pair of SM fermions for the complex scalar; see Appendix~\ref{colored scalar section}. In the last row, $F$ stands for $T$ or $B$.}
\label{table of fields}
\end{table}

\subsection{Tree-level decay}\label{tree decay section}
The scalar $s$ may dominantly decay into di-Higgs at the tree level due to the coupling~\eqref{shh coupling}:
\al{
\Gamma\fn{s\to hh}
	& 
	=	{\mu_\tx{eff}^2\over32\pi m_s}\sqrt{1-{4m_h^2\over m_s^2}}\sin^2\theta.
	\label{s_to_hh}
}
For $m_s>2m_Z$, the partial decay rate into the pair of vector bosons $s\to VV$ with $V=W,Z$ are
\al{
\Gamma\fn{s\to VV}
	&=	{m_s^3\over32\pi v^2}\delta_V\sqrt{1-4x_V}\paren{1-4x_V+12x_V^2}\,\sin^2\theta,
}
where $\delta_Z=1$, $\delta_W=2$, and $x_V=m_V^2/m_s^2$; see e.g.\ Ref.~\cite{Djouadi:2005gi}.
Similarly for $m_s>2m_t$, the partial decay width into a top-quark pair is
\al{
\Gamma\fn{s\to t\bar t}
	&=	{N_cm_sm_t^2\over8\pi v^2}\paren{1-{4m_t^2\over m_s^2}}^{3/2}\sin^2\theta.
}
Note that the tree-level branching ratios become independent of $\theta$ thanks to the parametrization~\eqref{shh coupling}.

The total decay width $\Gamma_\tx{total}$ is the sum of the above rates at the tree level.
In the small mixing limit $\theta^2\ll1$, the tree-level decay width becomes small and the loop level decay, which is described in Sec.~\ref{loop decay}, can be comparable to it.
The diphoton constraint is severe in this parameter region, as will be discussed in Sec.~\ref{constraint}.

\begin{figure}[tp]
\centering
\includegraphics[width=0.4\textwidth]{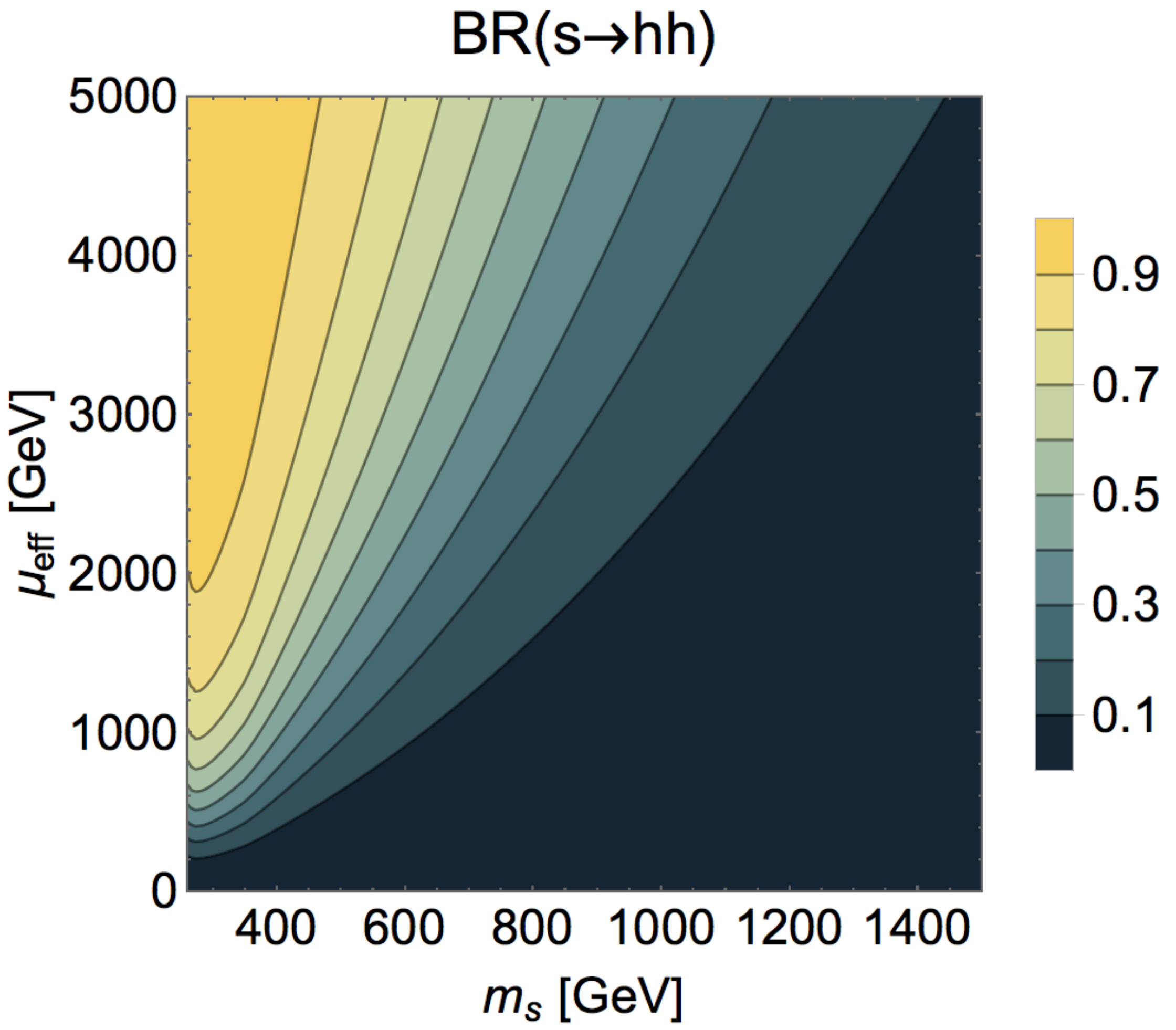}
\includegraphics[width=0.4\textwidth]{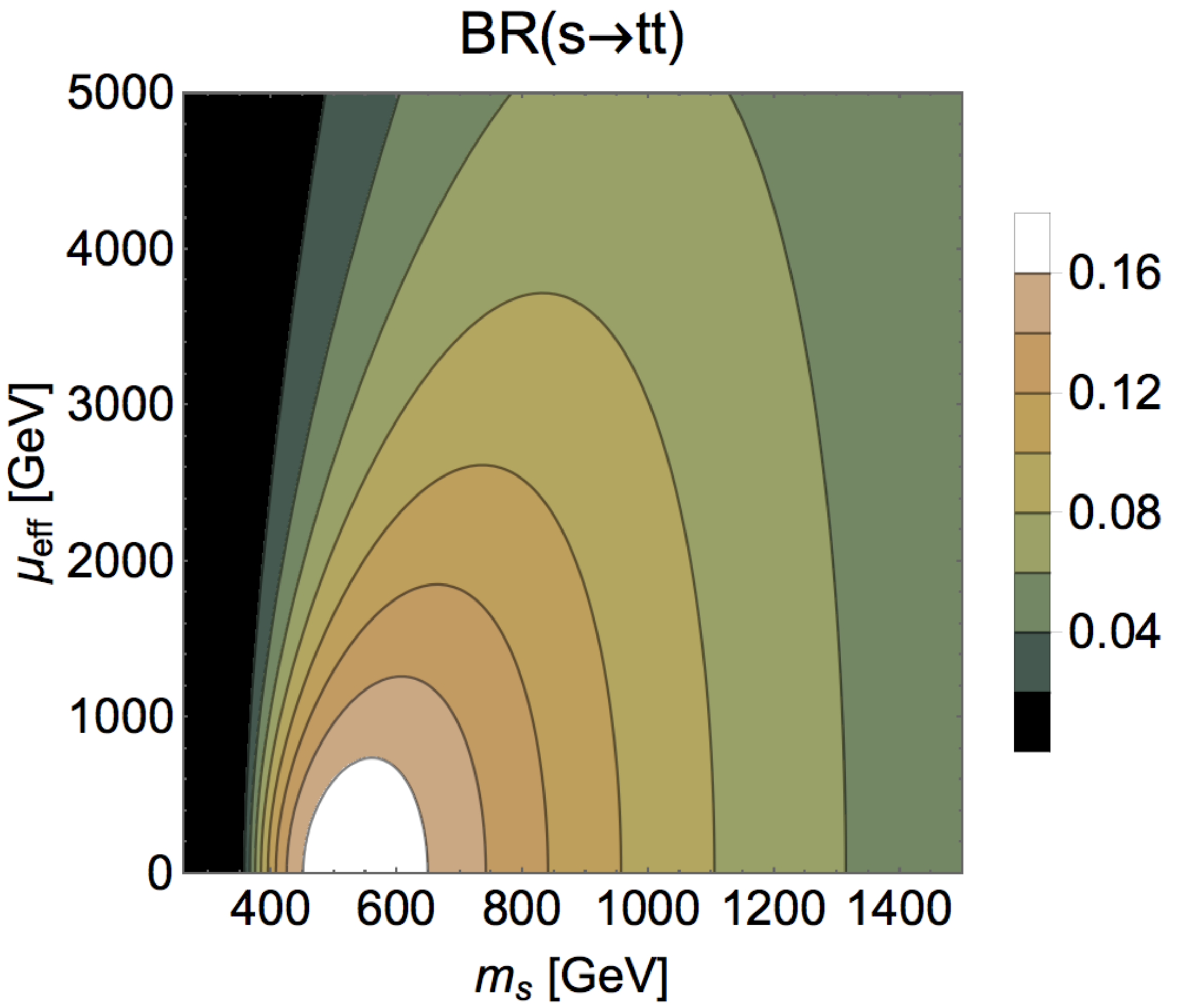}\smallskip\\
\includegraphics[width=0.4\textwidth]{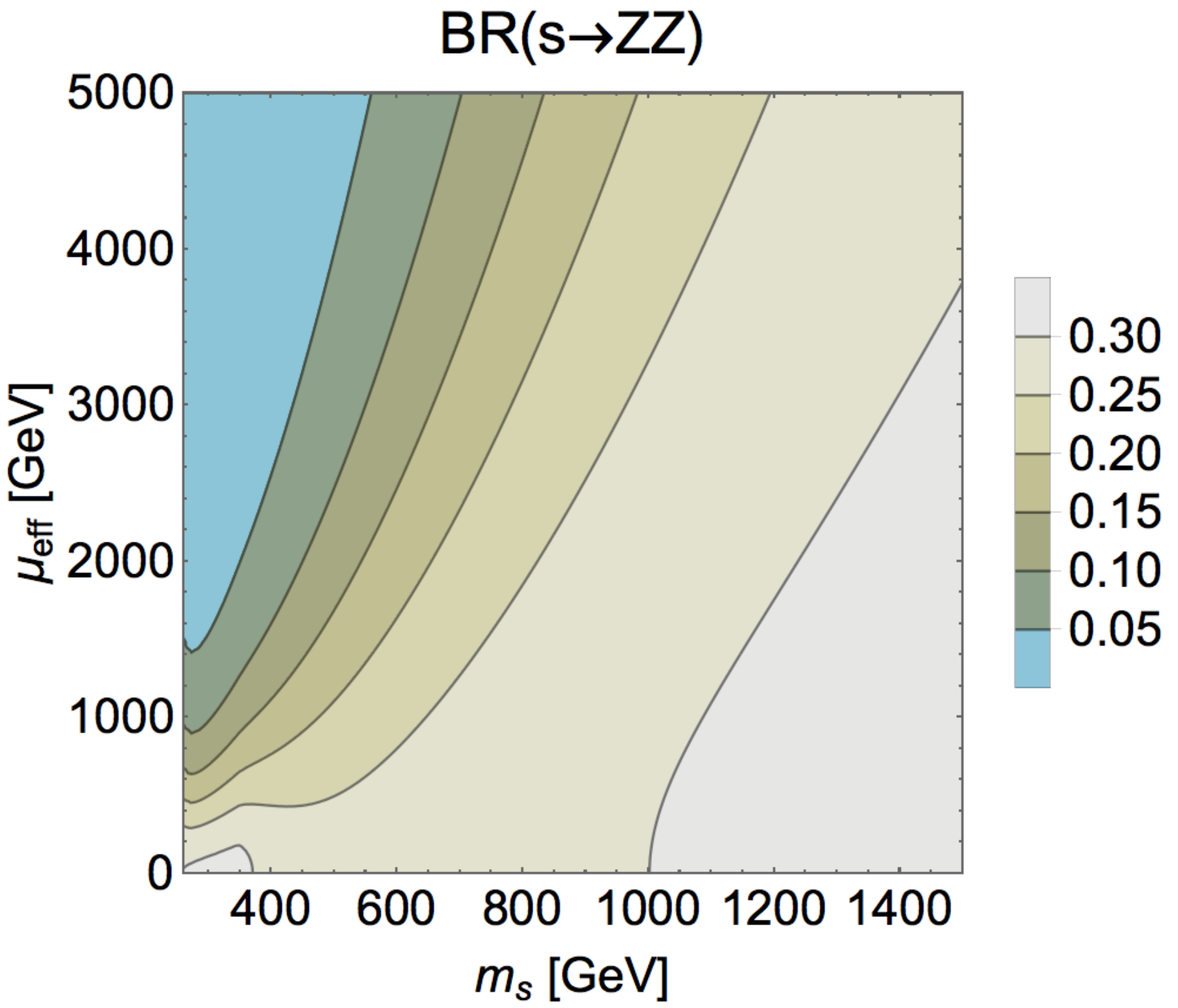}
\includegraphics[width=0.4\textwidth]{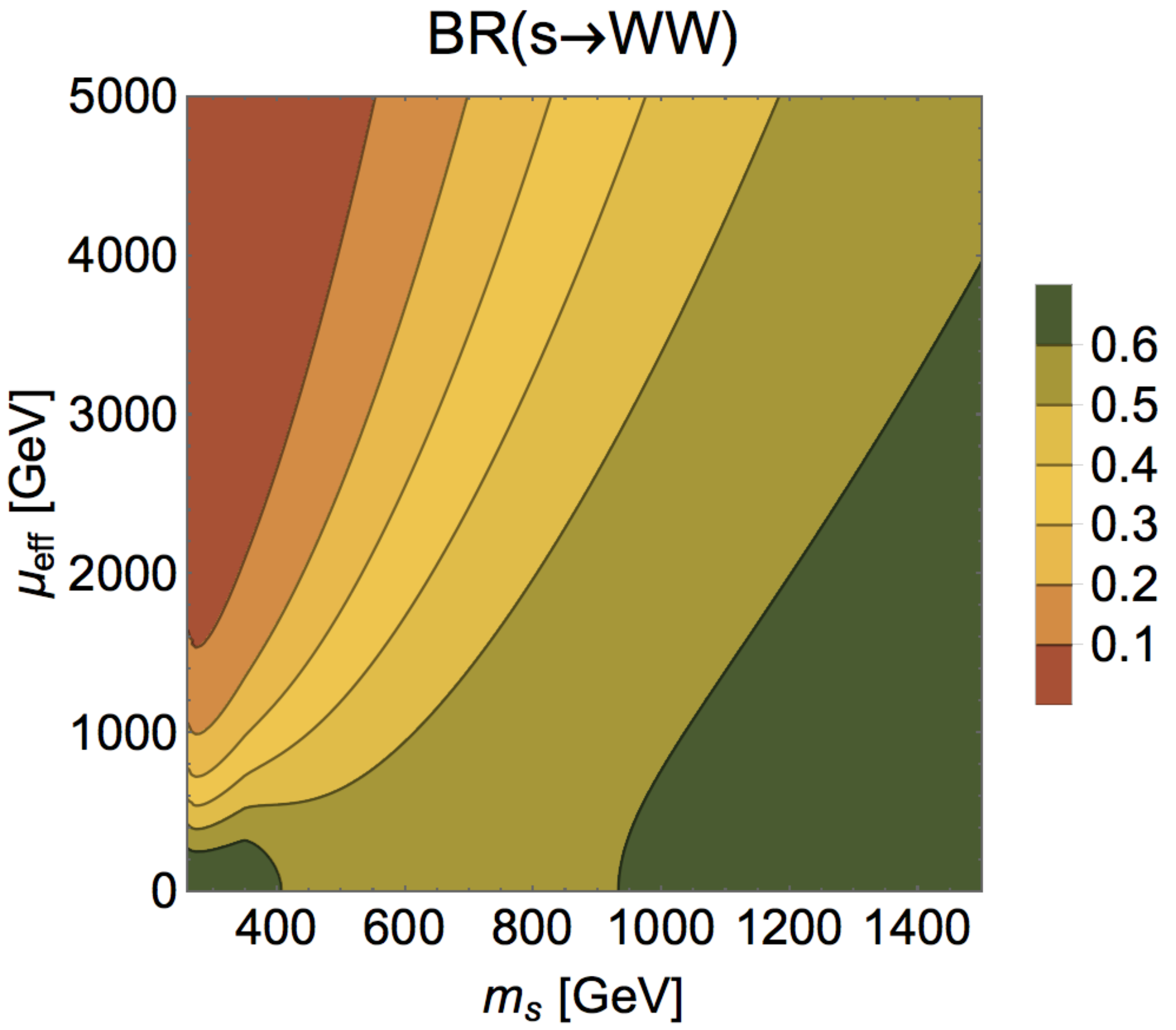}
\caption{Tree-level branching ratio for the decay of $s$ in the $\mu_\tx{eff}$ vs $m_s$ plane.}\label{BR figure}
  \end{figure}

In Fig.~\ref{BR figure}, we plot the tree-level branching ratios in the $\mu_\tx{eff}$ vs $m_s$ plane. Note that the $\theta$-dependence drops out of the tree-level branching ratios when we use $\mu_\tx{eff}$ as a phenomenological input parameter as in Eq.~\eqref{shh coupling} because then all the decay channels have the same $\theta$ dependence $\propto\sin^2\theta$.

\subsection{Effective coupling to photons and gluons}
We first consider the vector-like top-partner $T$ as the colored particle running in the loop that is represented as the blob in Fig.~\ref{Di-Higgs figure}.
The bottom-partner $B$ can be treated in the same manner, as well as the colored scalars.

The mass of the top partner is given as
\al{
M_T	&=	m_T+y_T f,
}
where $m_T$ and $y_T$ are the vector-like mass of $T$ and the Yukawa coupling between $T$ and $S$, respectively.
The top-partner $T$ mixes with the SM top quark.
We note that limit $m_T\to 0$ corresponds to an effective dilaton model.\footnote{
The particular dilaton model in Ref.~\cite{Abe:2012eu} corresponds to the identification of the lighter 125\,GeV scalar to be an $S$-like one, contrary to this paper.
}

Given the kinetic term of gluon that is non-canonically normalized,
\al{
\mathcal L_\tx{eff}
	&=	-{1\over 4g_s^2}G_{\mu\nu}^aG^{a\mu\nu},
}
the effective coupling after integrating out the top and $T$ can be obtained by the replacement $\left\langle S\right\rangle\to S$ and $\left\langle H^0\right\rangle\to H^0$ in the running coupling; see e.g.\ Refs.~\cite{Carena:2012xa,Abe:2012eu}:
\al{
{1\over g_s^2}
	&\longrightarrow
		{1\over g_s^2}-{2\over\paren{4\pi}^2}\paren{ b_g^\tx{top}{h\cos\theta+ s\sin\theta\over v}+\Delta b_g\,y_T{-h\sin\theta+ s\cos\theta\over M_T}},
		\label{replacement}
}
where $b_g^\tx{top}$ and $\Delta b_g$ are the contributions of top and $T$ to the beta function, respectively.
To use this formula, we need to assume the new colored particles are slightly heavier than the neutral scalar.
For a Dirac spinor in the fundamental representation, $b_g^\tx{top}=\Delta b_g={1\over2}\times{4\over3}={2\ov3}$.
The resultant effective interactions for the canonically normalized gauge fields are
\al{
\mathcal L_\tx{eff}^{hgg}
	&=
		{\alpha_s\over8\pi v}\paren{
			 b_g^\text{top} \cos\theta - \Delta b_g\eta\sin\theta
			}
		h\,G_{\mu\nu}^aG^{a\mu\nu},
			\label{hgg effective}\\
\mathcal L_\tx{eff}^{ s gg}
	&=
		{\alpha_s\over8\pi v}\paren{
			\Delta b_g\eta\cos\theta +  b_g^\text{top}\sin\theta
			}
		 s\,G_{\mu\nu}^aG^{a\mu\nu},\\
\mathcal L_\tx{eff}^{h\gamma\gamma}
	&=
		{\alpha\over8\pi v}\paren{
			b_\gamma^\text{SM}\cos\theta-\Delta b_\gamma\,\eta\sin\theta
			}
		hF_{\mu\nu}F^{\mu\nu},\\
\mathcal L_\tx{eff}^{ s\gamma\gamma}
	&=
		{\alpha\over8\pi v}\paren{
			\Delta b_\gamma\,\eta\cos\theta+b_\gamma^\text{SM}\sin\theta
			}
		 s F_{\mu\nu}F^{\mu\nu},
			\label{s gam gam eff}
}
where $F_{\mu\nu}$ being the (canonically normalized) field strength tensor of the photon, $\alpha_s$ and $\alpha$ denoting the chromodynamic and electromagnetic fine structure constants, respectively,
$N_c=3$, $b_\gamma^\text{SM}\simeq-6.5$ and
\al{
\eta
	&=	y_TN_T{v\over M_T},
	\label{EqEta}
}
with $N_T$ being the number of $T$ introduced. The values $\Delta b_g={1\ov2}\times{4\ov3}={2\ov3}$ and $\Delta b_\gamma=N_cQ_T^2\times{4\ov3}={16\ov9}$ are listed in Table~\ref{table of fields}.

The bottom partner $B$ can be treated exactly the same way. According to Table~1, $\Delta b_\gamma$ becomes one fourth compared to the above.

For the colored scalar $\phi$, its diagonal mass is given as
\al{
M_\phi^2
	&=	m_\phi^2+{\kappa_\phi\over2} \left\langle S\right\rangle^2,
}
where we have assumed the $Z_2$ symmetry $S\to-S$ for simplicity;
$m_\phi$ is the original diagonal mass in the Lagrangian; and 
$\kappa_\phi$ is the quartic coupling between $S$ and $\phi$.\footnote{
 The three point interaction between the neutral and the colored scalar can be introduced.
 If the sign of the three and the four point couplings are opposite, $\eta$ can be enhanced in some parameter region.
}
The possible values of the electromagnetic of $\phi$ are
$Q=-1/3$ and $-4/3$ for the leptoquark $\phi_{\bs 3}$;
$Q=1/3$, $-2/3$, and $4/3$ for the color-sextet $\phi_{\bs 6}$; and
$Q=0$ and $-1$ for the color-octet $\phi_{\bs 8}$; see Appendix~\ref{colored scalar section}.
Correspondingly the values of $\Delta b_g$ are ${1\over2}\times{1\over3}={1\ov6}$, $\paren{{N_c\ov2}+1}\times{1\ov3}={5\ov6}$, and $N_c\times{1\ov3}=1$, and $\Delta b_\gamma$ are $Q^2$, $2Q^2$, and ${8\ov3}Q^2$.
Again the effective interactions are obtained as in Eqs.~\eqref{hgg effective}--\eqref{s gam gam eff} from the replacement~\eqref{replacement} with the substitution $y_T/M_T\to \kappa_\phi f/M_\phi^2$, where $f$ has been the VEV of $S$; see Eq.~\eqref{VEVs}.
Note that the expression for $\eta$ is now $\eta=\kappa_\phi N_\phi{fv/M_\phi^2}$, where $N_\phi$ is the number of $\phi$ introduced.
We list all these parameters in Table~\ref{table of fields}.

\subsection{Loop-level decay}\label{loop decay}
No direct contact to the gauge bosons are allowed for the singlet scalar $S$, and the tree-level decay of $s$ into a pair of gauge bosons is only via the mixing with the SM Higgs boson. Therefore the decay of $s$ to $gg$ and $\gamma\gamma$ are only radiatively generated. 
Given the effective operators from the loop of heavy colored particle
\al{
\mc L_\tx{eff}
	&=	-{\alpha_s b_g\over4\pi v} s G^a_{\mu\nu}G^{a\mu\nu}
		-{\alpha b_\gamma\over4\pi v} s F_{\mu\nu}F^{\mu\nu},
}
the partial decay widths are
\al{
\Gamma\fn{ s\to gg}
	&=	\paren{\alpha_s b_g\over4\pi v}^2{2m_ s^3\over\pi},	&
\Gamma\fn{ s\to\gamma\gamma}
	&=	\paren{\alpha b_\gamma\over4\pi v}^2{m_ s^3\over4\pi},
}
where the factor 8 difference comes from the number of degrees of freedom of gluons in the final state.
Concretely,
\al{
b_g
	&=	-{1\over2}\paren{\Delta b_g\,\eta\cos\theta+ b_g^\tx{top}\sin\theta}, \label{bg given}\\
b_\gamma
	&=	-{1\ov2}\paren{\Delta b_\gamma\,\eta\cos\theta+b_\gamma^\text{SM}\sin\theta}.
}
If we go beyond the scope of this paper and allow the particles in the loop to be charged under $SU(2)_L$, then the loop contribution to the decay channels to $Z\gamma$, $ZZ$ and $W^+W^-$ might also become significant; see e.g.\ Ref.~\cite{Kim:2015vba}.

\section{Production of singlet scalar at hadron colliders}\label{signal section}

We calculate the production cross section of $s$ via the gluon fusion with the narrow width approximation:\footnote{
The colored particles running in the blob in Fig.~\ref{Di-Higgs figure} might also have a direct coupling with the quarks in the proton, and possibly change the production cross section of $s$ if it is extremely large. In this paper we assume that this is not the case.
}
\al{
\h\sigma\fn{gg\to s}
	&=	{\pi^2\over8m_ s}\Gamma\fn{ s\to gg}\delta\fn{\h\sigma-m_ s^2}
	=	 \sigma_s m_ s^2\delta\fn{\h\sigma-m_ s^2},
		\label{narrow width limit}
}
where
\al{
 \sigma_s
	&:=	{\pi^2\over8m_ s^3}\Gamma\fn{ s\to gg}
	=	\paren{\alpha_s b_g\over4\pi v}^2{\pi\over4}
	=	36.5\,\tx{fb}\times\sqbr{b_g\over-1/3}^2\sqbr{\alpha_s\over0.1}^2.
		\label{sigma_s}
}
Therefore, we reach the expression with the gluon parton distribution function~(PDF) for the proton $g(x,\mu_F)$:
\al{
\sigma\fn{pp\to s}
	&=	 \sigma_s m_ s^2\int_0^1\df x_1\int_0^1\df x_2\,g\fn{x_1,\mu_F}\,g\fn{x_2,\mu_F}\,\delta\fn{x_1x_2s-m_ s^2} 
	=	 \sigma_s\tau{\df\mc L^{gg}\over\df\tau},
		\label{singlet production}
}
where $\tau:=m_s^2/s$ and
\al{
{\df\mc L^{gg}\over\df\tau}
	&=	\int_\tau^1{\df x\over x}g\fn{x,\mu_F}g\fn{\tau/x,\mu_F}
	=	\int_{\ln\sqrt{\tau}}^{\ln{1\over\sqrt{\tau}}}\df y\,
		g\fn{\sqrt{\tau}e^y,\sqrt{\tau s}}\,g\fn{\sqrt{\tau}e^{-y},\sqrt{\tau s}},
}
is the luminosity function, in which the factorization scale $\mu_F$ is taken to be $\mu_F=\sqrt{\tau s}$.\footnote{
Notational abuse of $s$ for the singlet scalar field and for the Mandelstam variable of $pp$ scattering should be understood.
}


\begin{figure}[tp]
\begin{center}
\includegraphics[width=0.7\textwidth]{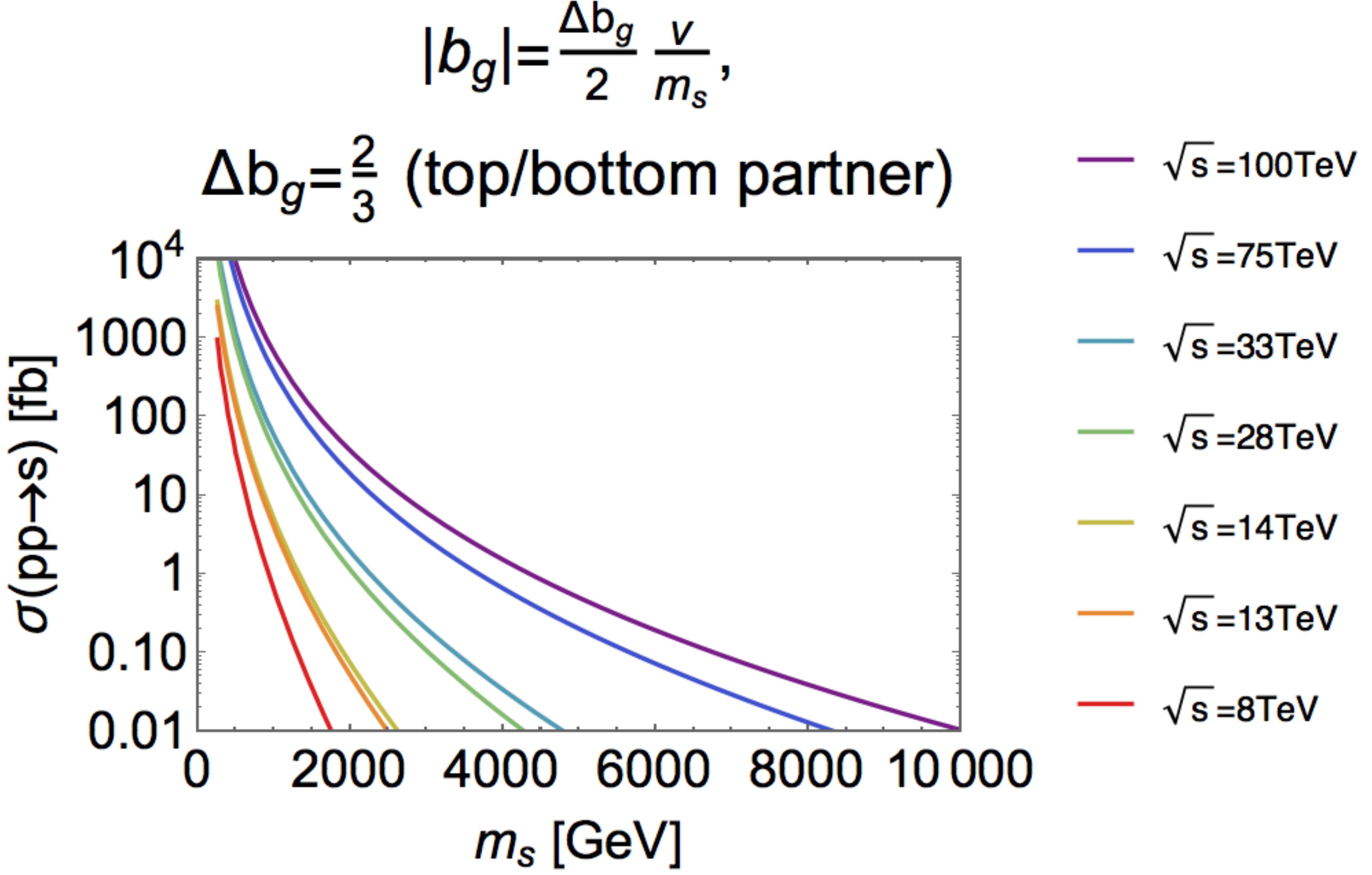}
  \end{center}
\caption{
  Production cross section $\sigma\fn{pp\to s}$ for $\ab{b_g}=\frac{\Delta b_g}{2}\frac{v}{m_s}$ with $\Delta b_g={2\ov3}$ (top/bottom partner).
	The result for other parameter can be obtained just by a simple scaling $\sigma\fn{pp\to s}\propto\paren{\Delta b_g}^2$;
	see Eq.~\eqref{sigma_s} with Eq.~\eqref{bg given} and Table~\ref{table of fields}.
	The $K$-factor is not included in this plot.
}\label{prospect}
  \end{figure}

Using the leading order CTEQ6L~\cite{Nadolsky:2008zw} PDF, we plot in Fig.~\ref{prospect} the production cross section $\sigma\fn{pp\to s}$ as a function of $m_s$ for a phenomenological benchmark setting $\ab{b_g}={\Delta b_g\ov2}{v\ov m_s}$ with $\Delta b_g={2\ov3}$ (top/bottom partner).
Other particles just scale as $\sigma\fn{pp\to s}\propto\paren{\Delta b_g}^2$.
The value $\sqrt{s}=14\TeV$ is motivated by the High-Luminosity LHC;
28\,TeV and 33\,TeV by the High-Energy LHC (HE-LHC); and
75\,TeV, and 100\,TeV by the Future Circular Collider (FCC)~\cite{FCC,Benedikt:1742294,Ball}.

We see that typically the top/bottom partner models give a cross section $\sigma\fn{pp\to s}\gtrsim1\fb$, which could be accessed by a luminosity of $\mc O\fn{\tx{ab}^{-1}}$, for the scalar mass $m_s\lesssim1.3\TeV$, 2\,TeV, and 4\,TeV at the LHC, HE-LHC, and FCC, respectively.

Several comments are in order:
\begin{itemize}
\item Our setting corresponds to putting $M_T = y_T N_T m_s$ in Eq.~\eqref{EqEta}
in order to reflect the naive scaling of $\eta\sim v/f$ with $f\sim m_s$; recall that we need $M_T\gtrsim m_s$ to justify integrating out the top partner to write down the effective interactions~\eqref{hgg effective}--\eqref{s gam gam eff}.
\item Here we have used the leading order parton distribution function. The higher order corrections may be approximated by multiplying an overall factor $K$, the so-called $K$-factor, which takes value $K\simeq1.6$ for the SM Higgs production at LHC; see e.g.\ Ref.~\cite{Djouadi:2005gi}.
\item \label{SM box} The SM cross section for $pp\to hh$ is of the order of 10\,fb and $10^3\fb$ for $\sqrt{s}=8\TeV$ and 100\,TeV, respectively~\cite{Baglio:2012np}. We are interested in the on-shell production of $s$, and the non-resonant SM background can be discriminated by kinematical cuts. The detailed study is beyond the scope of this paper and will be presented elsewhere.

\item When we consider the new resonance with a narrow width~\eqref{narrow width limit}, we can neglect the box contribution from the extra colored particles as the box contribution gets a suppression factor\footnote{
In the SM, the $gg\to hh$ cross section takes the following form at the leading order~\cite{Baglio:2012np}:
\als{
\h\sigma^\tx{SM}_\tx{LO}\fn{gg\to hh}
	&=	\int_{\h t_-}^{\h t_+}\df\h t{G_\tx{F}^2\alpha_s^2\over256\paren{2\pi}^3}\sqbr{
			\ab{{\mu_{hhh}v\ov\paren{\h s-m_h^2}+im_h\Gamma_h}F^\tx{SM}_\bigtriangleup+F^\tx{SM}_\Box}^2+\ab{G^\tx{SM}_\Box}^2
			},
}
where $G_\tx{F}$ is the Fermi constant; $\mu_{hhh}=3m_h^2/v$ is the $hhh$ coupling in the SM; and $F^\tx{SM}_\bigtriangleup$, $F^\tx{SM}_\Box$, and $G^\tx{SM}_\Box$ are the triangular and box form factors, approaching $F^\tx{SM}_\bigtriangleup\to2/3$, $F^\tx{SM}_\Box\to-2/3$, and $G^\tx{SM}_\Box\to0$  in the large top-quark-mass limit. A large cancellation takes place between $F^\tx{SM}_\bigtriangleup$ and $F^\tx{SM}_\Box$ as is well known.

 For the \emph{on-shell} resonance production of $s$, on the other hand, the triangle contribution from the fermion loops dominates over the box loop contribution: The new triangle contribution for $s$ can be well approximated by replacing the expression for the SM as
\als{
\mu_{hhh}
	&\to \mu_{\rm eff}\sin\theta, &
\,m_h
	&\to	m_s, &
\Gamma_h 
	&\to	\Gamma_s, &
F^\tx{SM}_\bigtriangleup 
	&\to	\Delta b_g \eta \cos \theta + b_g^{\rm top} \sin \theta, 
}
and the new box contribution of the top partner can be obtained from that of the SM-top quark with the multiplicative factor
\als{
{N_Ty_T^2\sin^2\theta\ov y_t^2/2}\,{y_T^2f^2\ov M_T^2}.
}
Finally, taking the ratio of the size of the box contribution and the triangle contribution with $\Delta b_g=2/3$ and $\eta=y_TN_Tv/M_T\sim N_Tv/M_T$, $y_T \sim y_t$, and 
$m_s \Gamma_s
	\sim	\mu_\tx{eff}^2\sin^2\theta/32\pi$, we get the result in Eq.~\eqref{box suppression}.
}	
\begin{align}
 {\mu_\tx{eff}M_T\ov32\pi v^2}\sin^3\theta
	&\sim	10^{-4}\sqbr{\mu_\tx{eff}\ov1\TeV}\sqbr{M_T\ov1\TeV}\sqbr{\sin\theta\ov0.1}^3 \ll 1.
		\label{box suppression}
\end{align}
\end{itemize}

\section{LHC constraints}\label{constraint}
We examine LHC constraints on the model for various $m_s$.
That is, we verify constraints from 125\,GeV Higgs signal strength, from $s\to ZZ\to4l$ search, from $s\to\gamma\gamma$ search, and from the direct search of the colored particles running in the blob in Fig.~\ref{Di-Higgs figure}.

\subsection{Bound from Higgs signal strength}

\begin{figure}[tp]
\begin{center}
\includegraphics[width=0.3\textwidth]{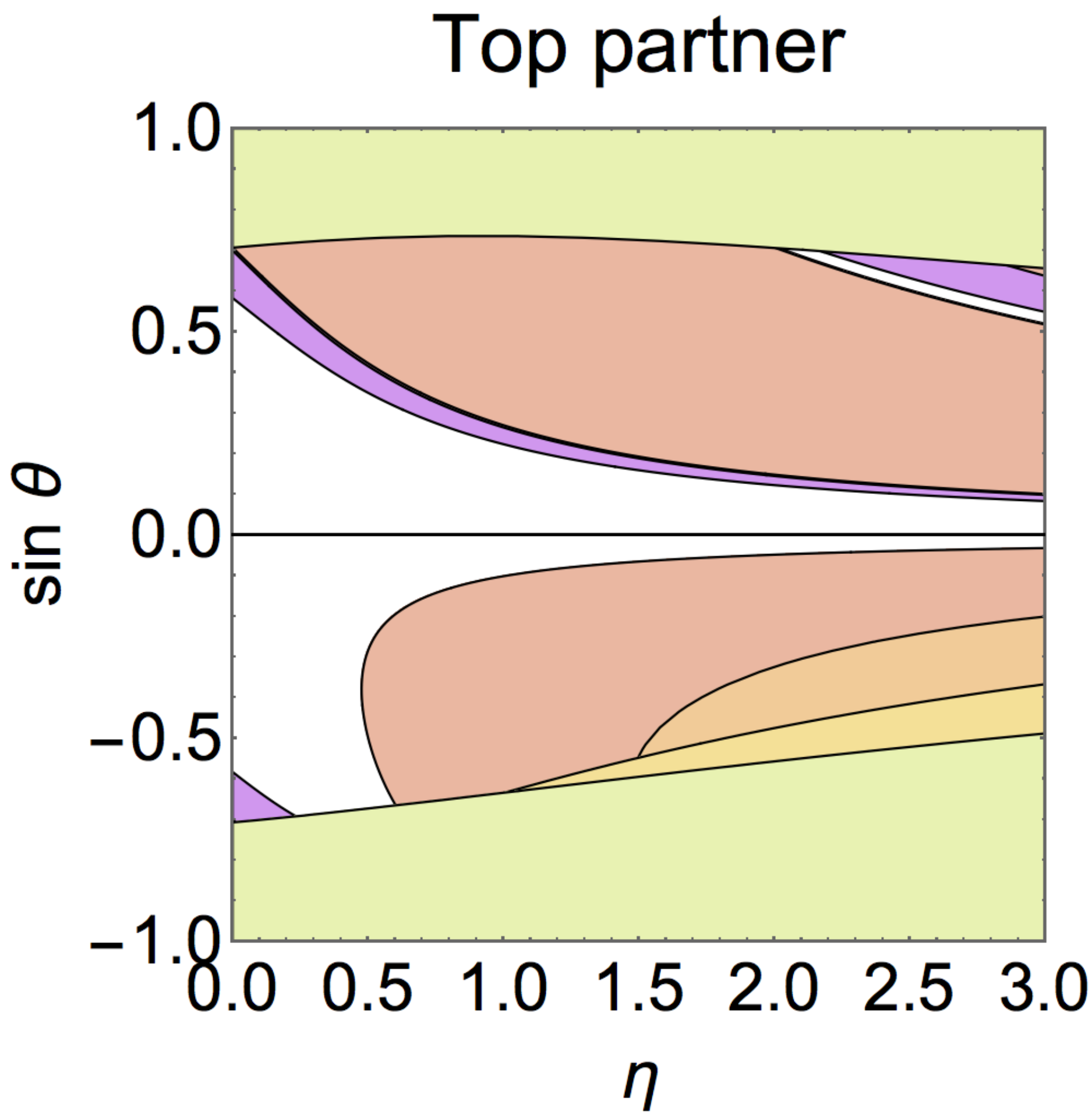}
\includegraphics[width=0.3\textwidth]{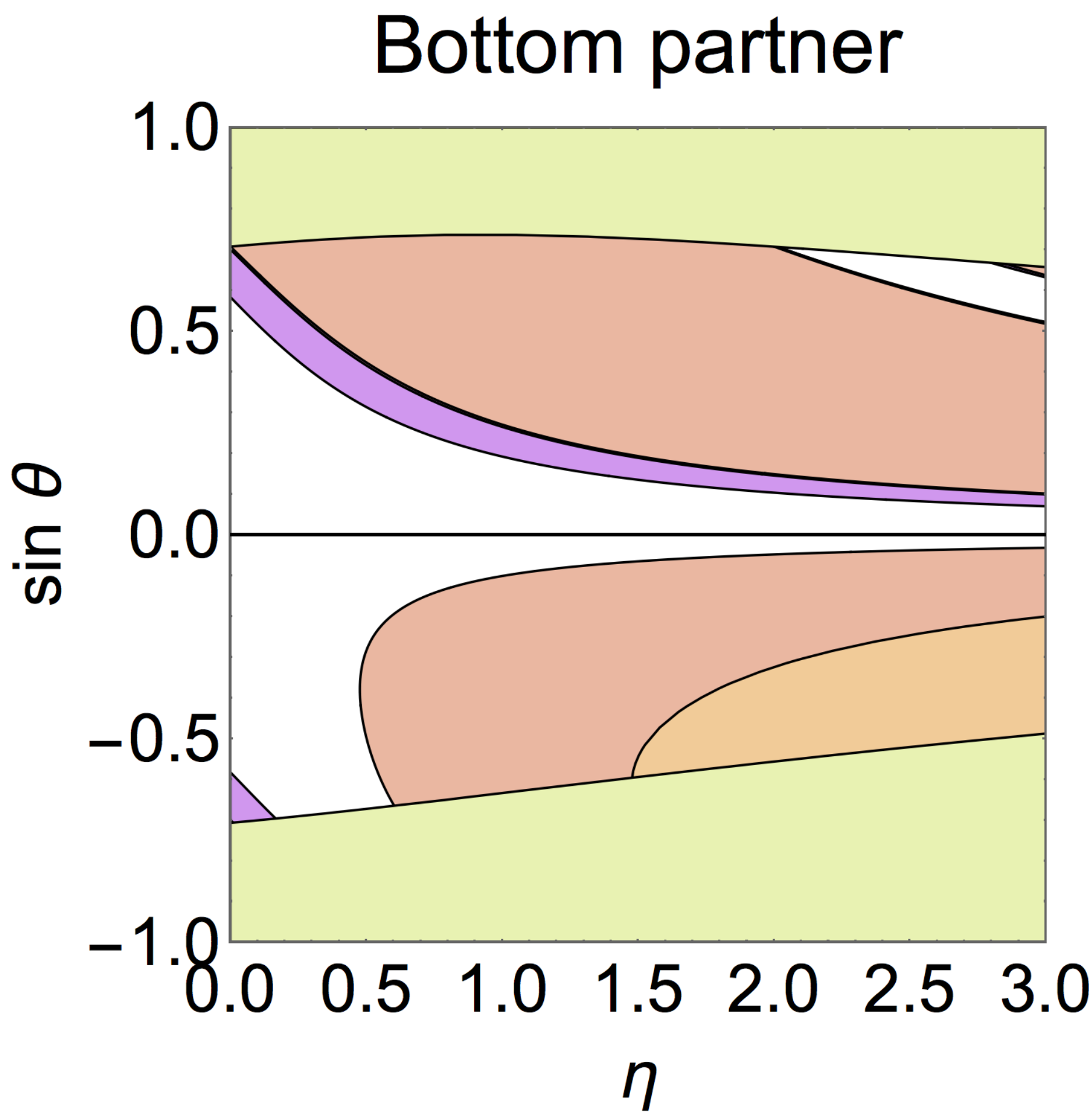}
\includegraphics[width=0.3\textwidth]{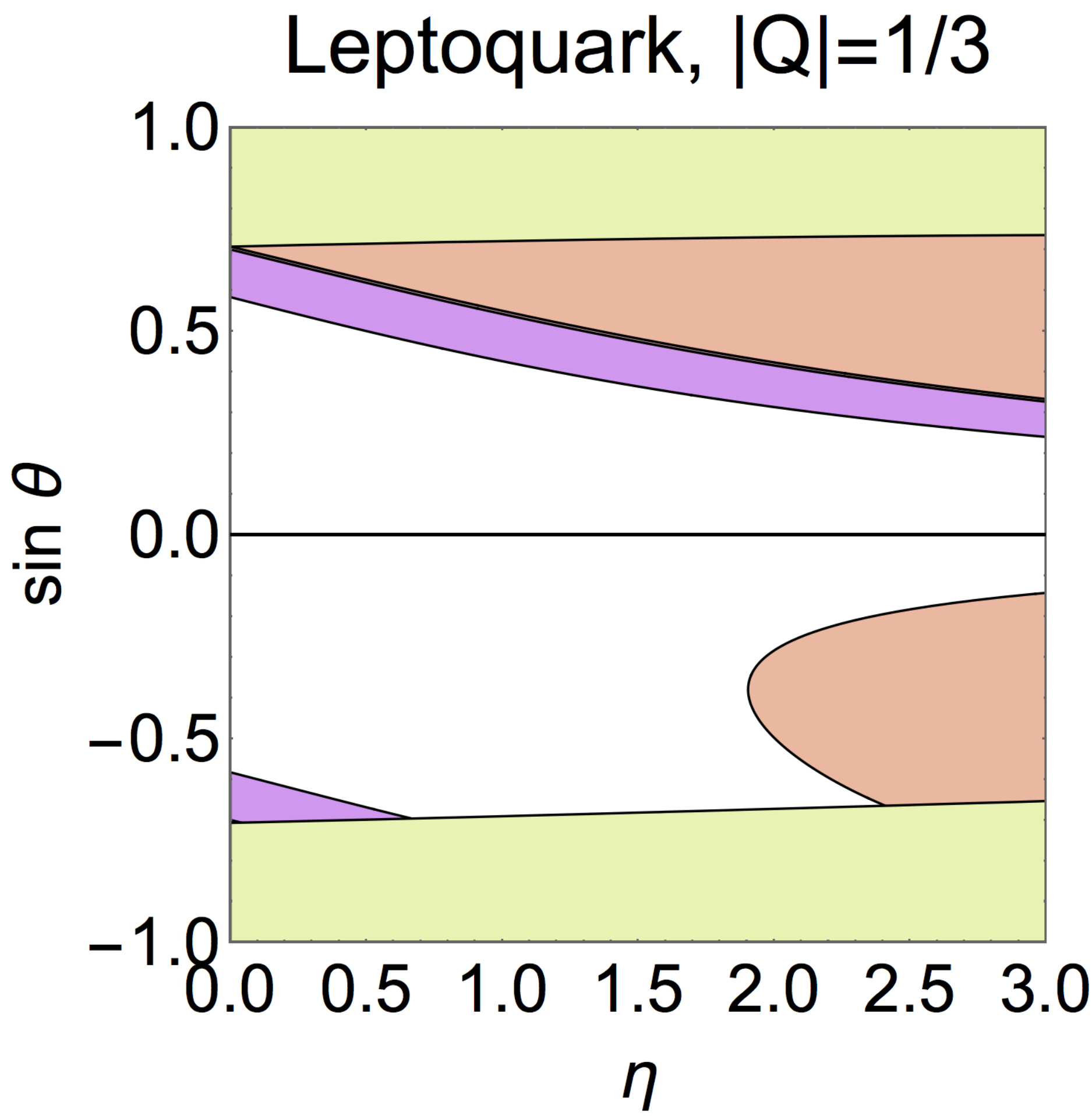}\medskip\\
\includegraphics[width=0.3\textwidth]{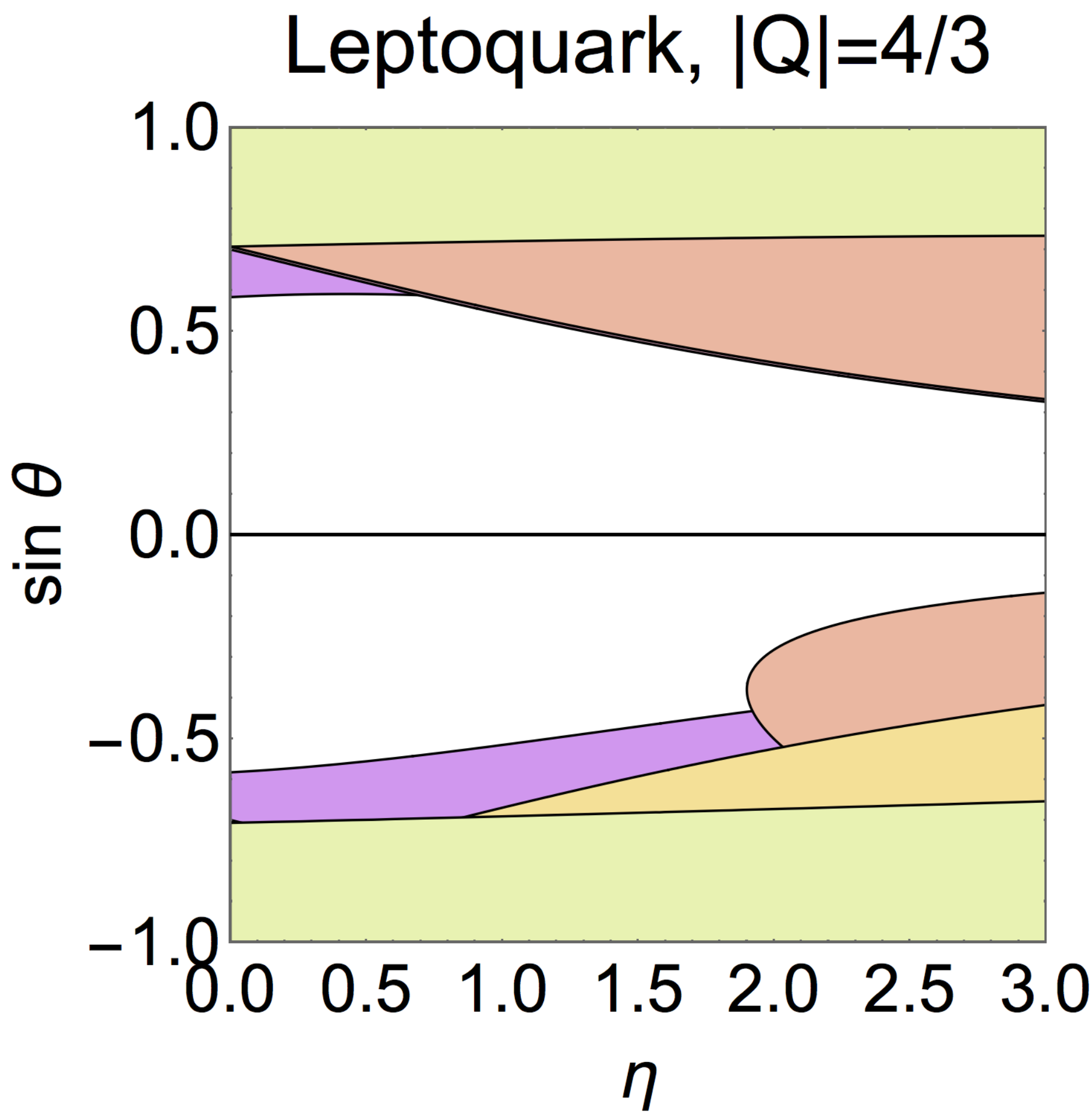}
\includegraphics[width=0.3\textwidth]{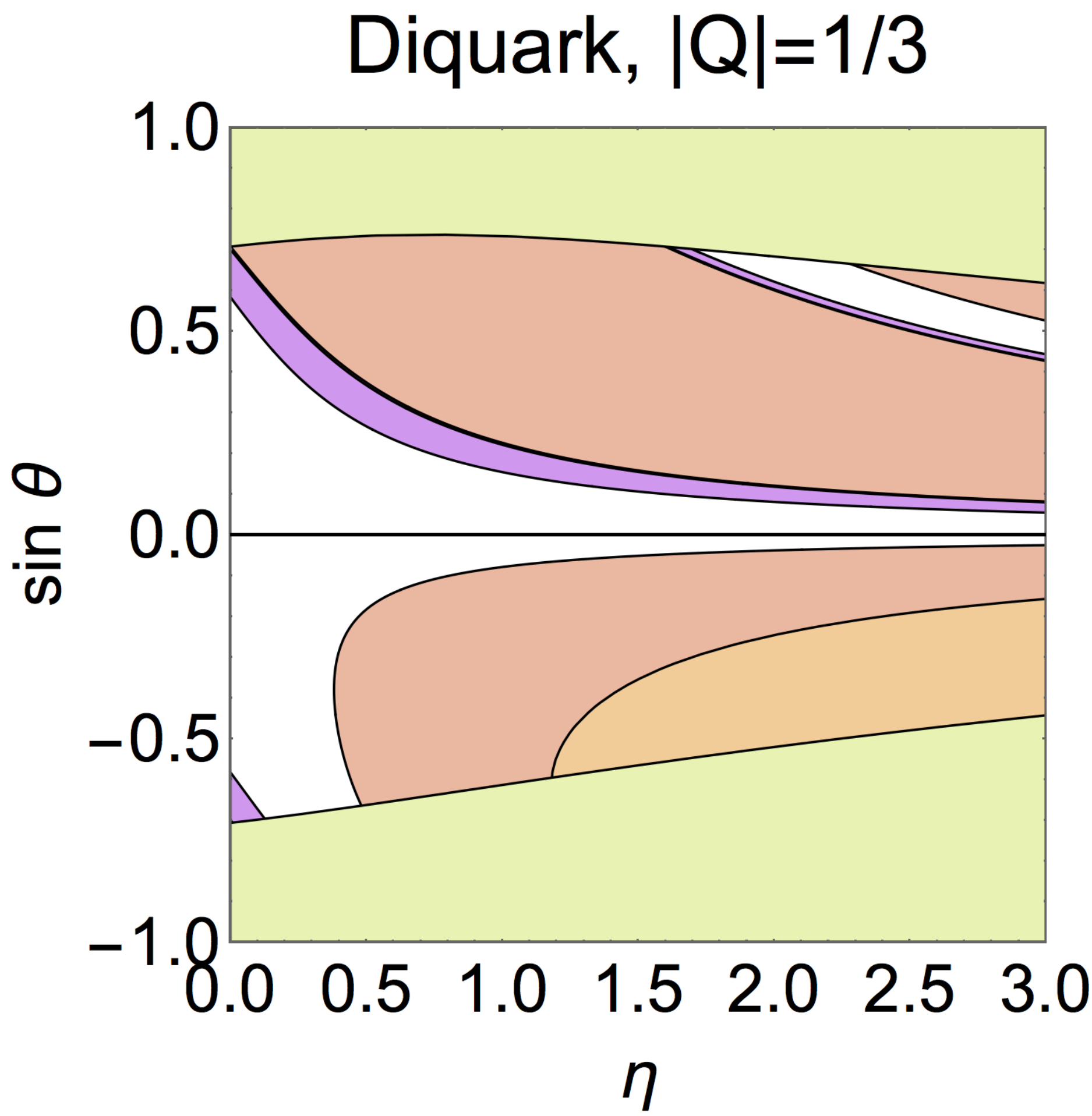}
\includegraphics[width=0.3\textwidth]{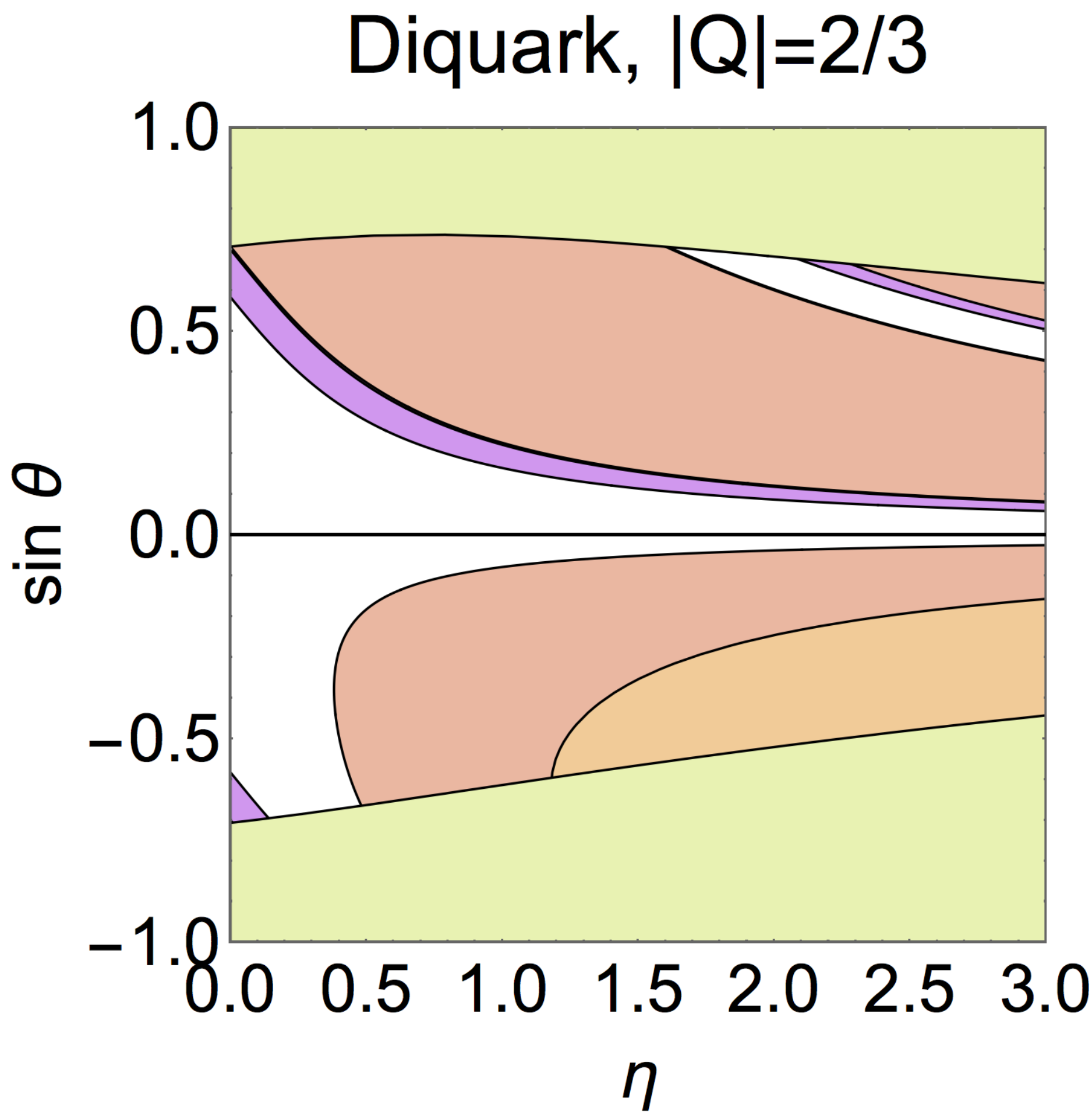}\medskip\\
\includegraphics[width=0.3\textwidth]{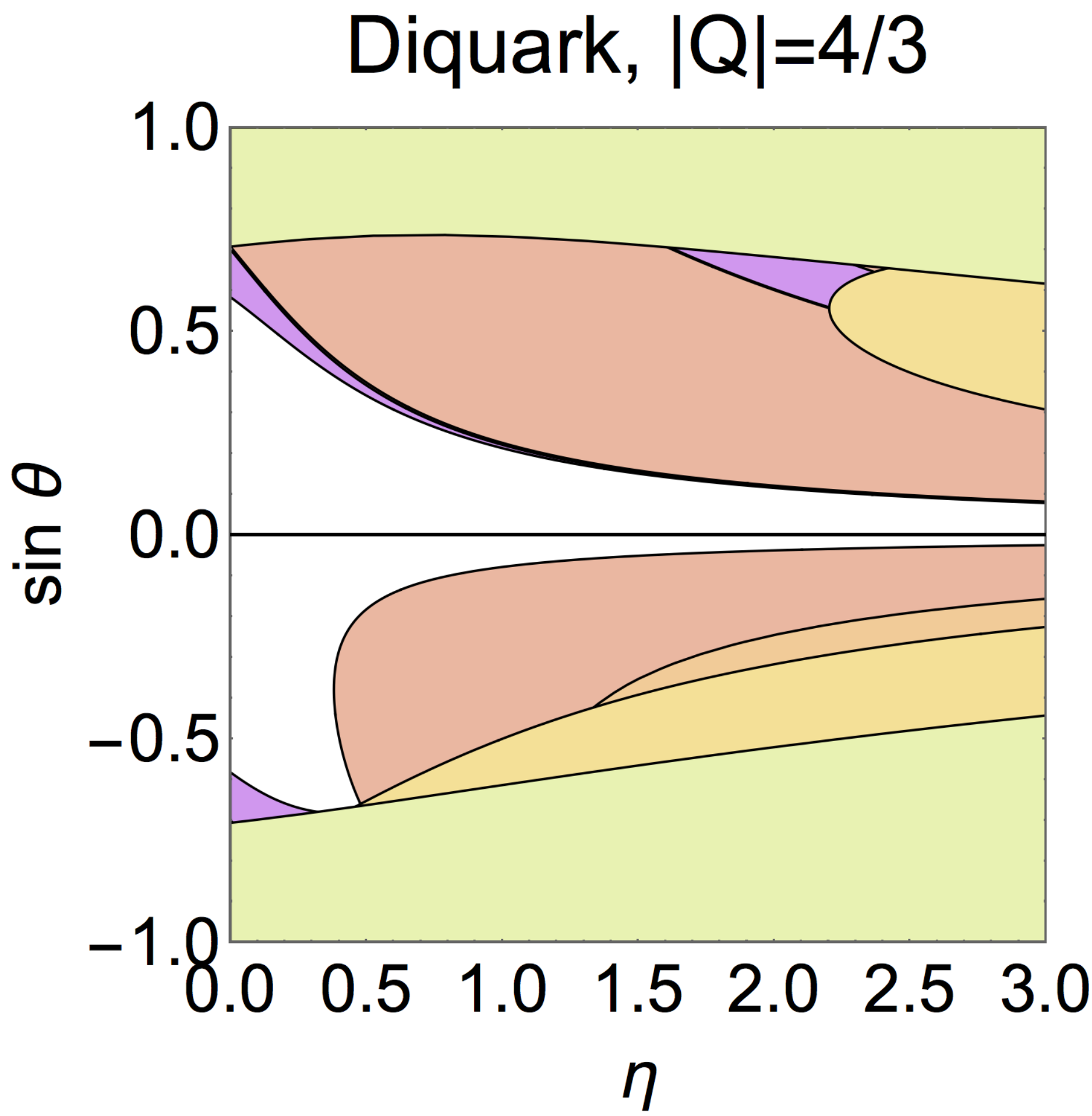}
\includegraphics[width=0.3\textwidth]{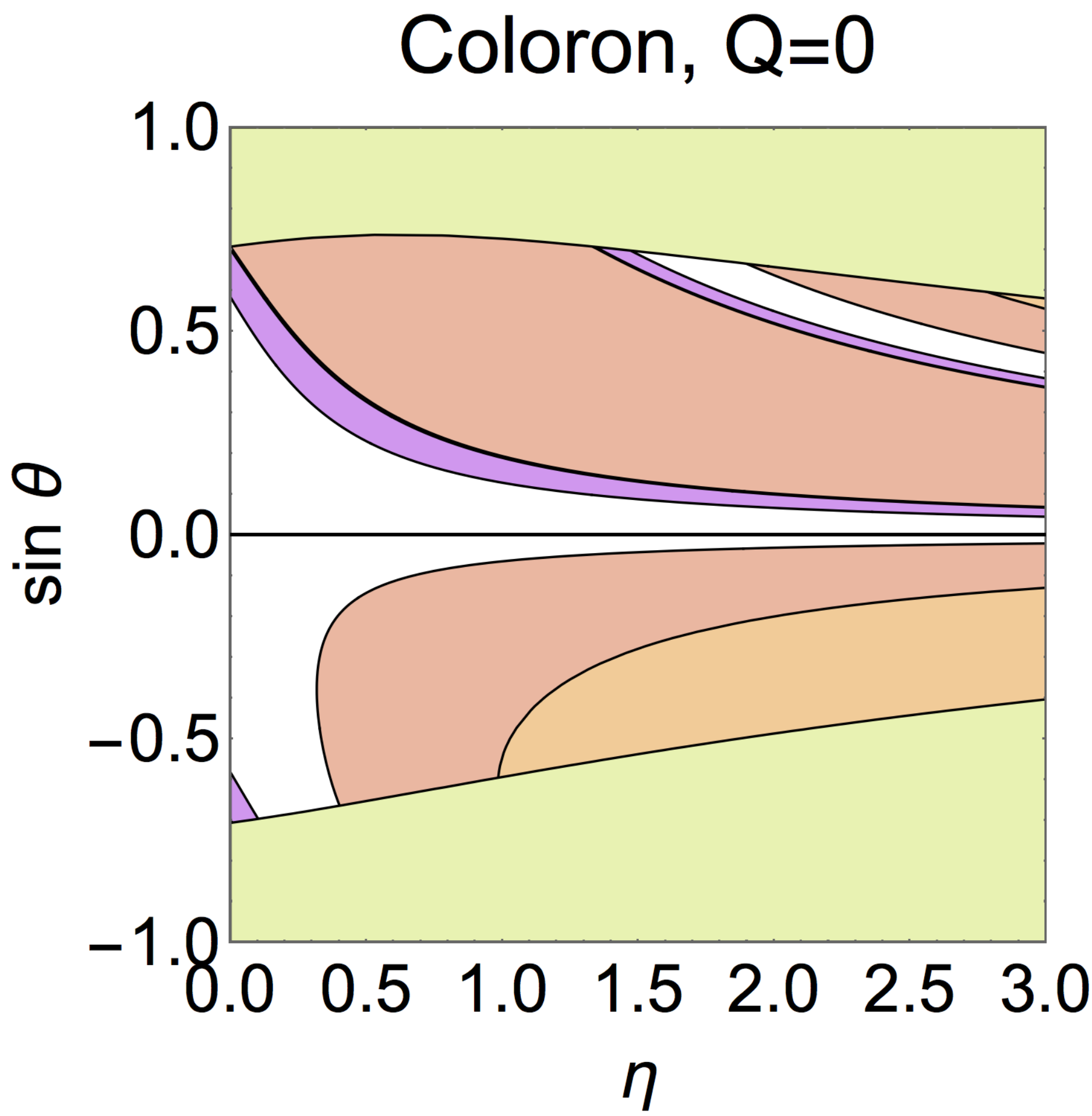}
\includegraphics[width=0.3\textwidth]{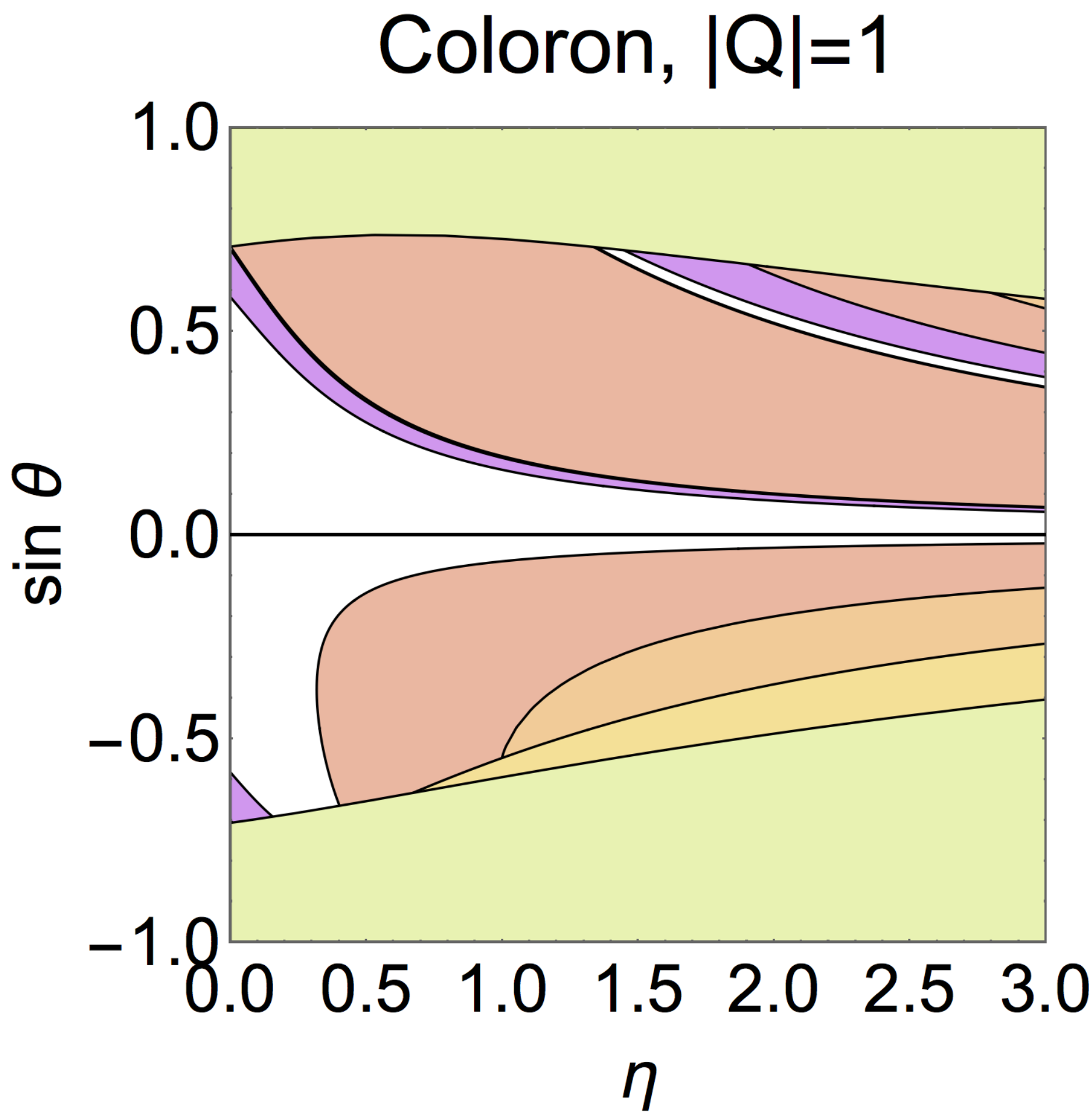}
  \end{center}
\caption{2$\sigma$-excluded regions from the signal strength of 125\,GeV Higgs are shaded.
The color represents the contribution from each channel; see Fig.~\ref{each} for details.}\label{combined}
  \end{figure}

\begin{figure}
\begin{center}
\includegraphics[width=0.24\textwidth]{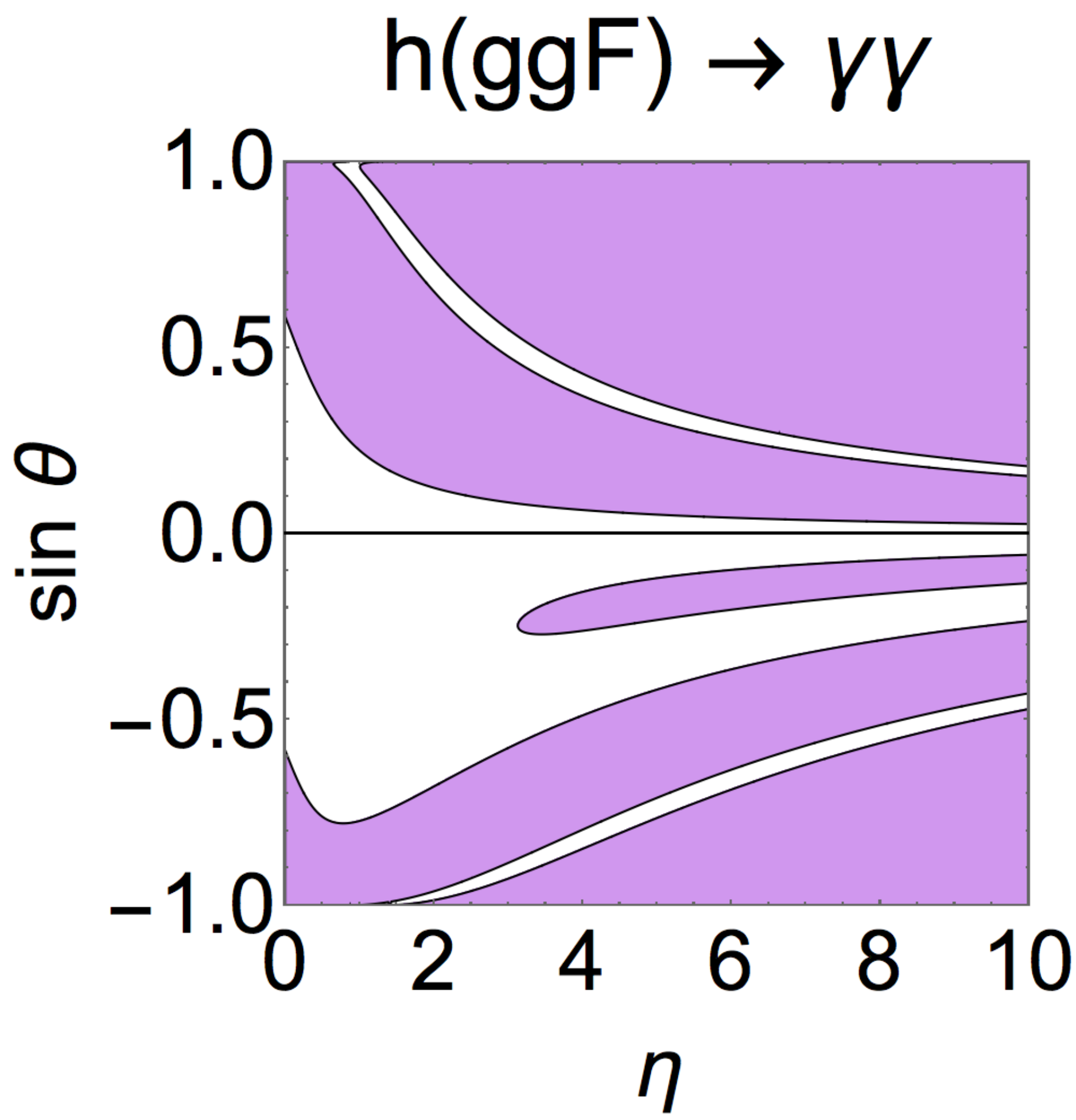} 
\includegraphics[width=0.24\textwidth]{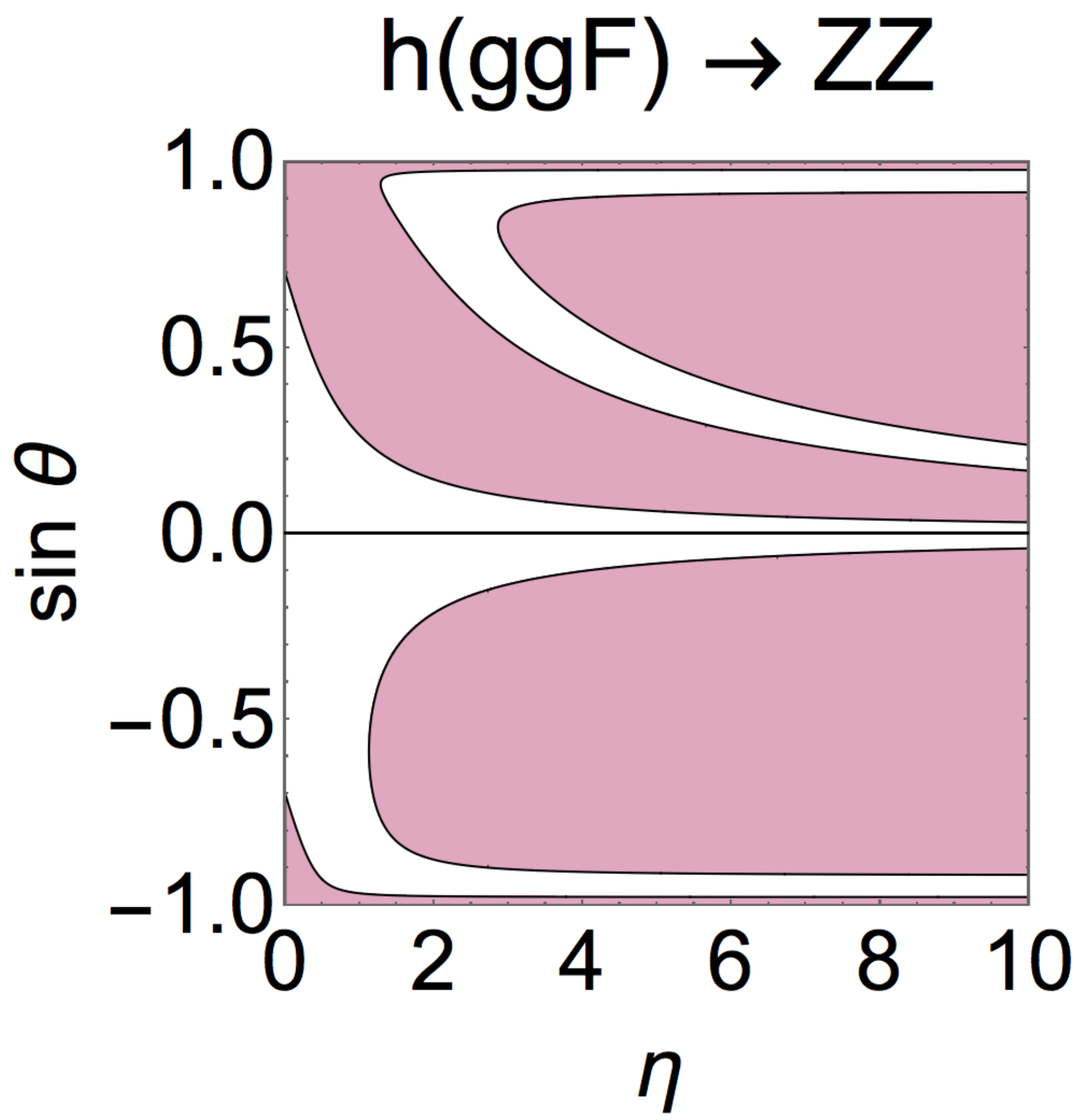} 
\includegraphics[width=0.24\textwidth]{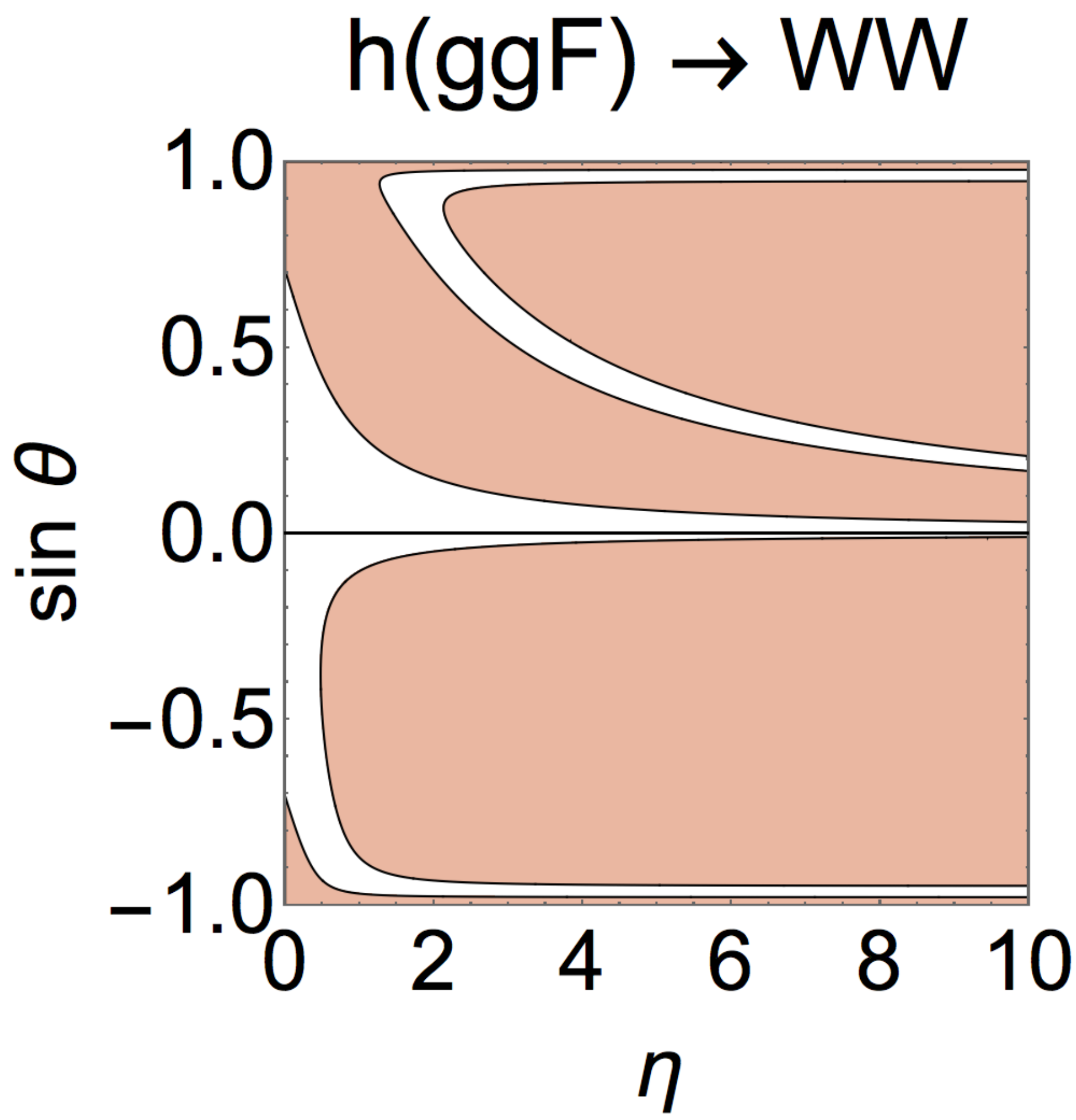} 
\includegraphics[width=0.24\textwidth]{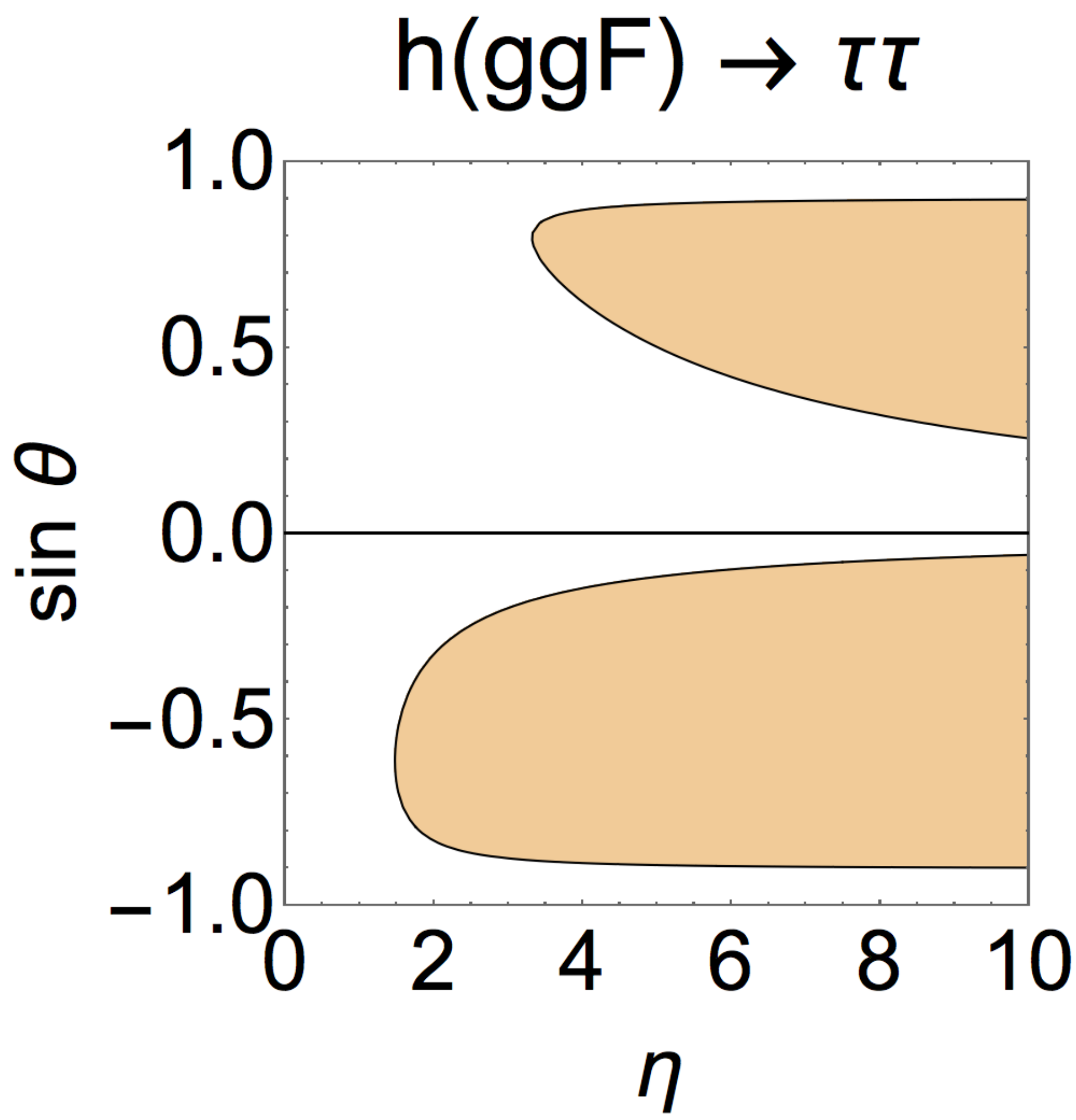}\\
\includegraphics[width=0.24\textwidth]{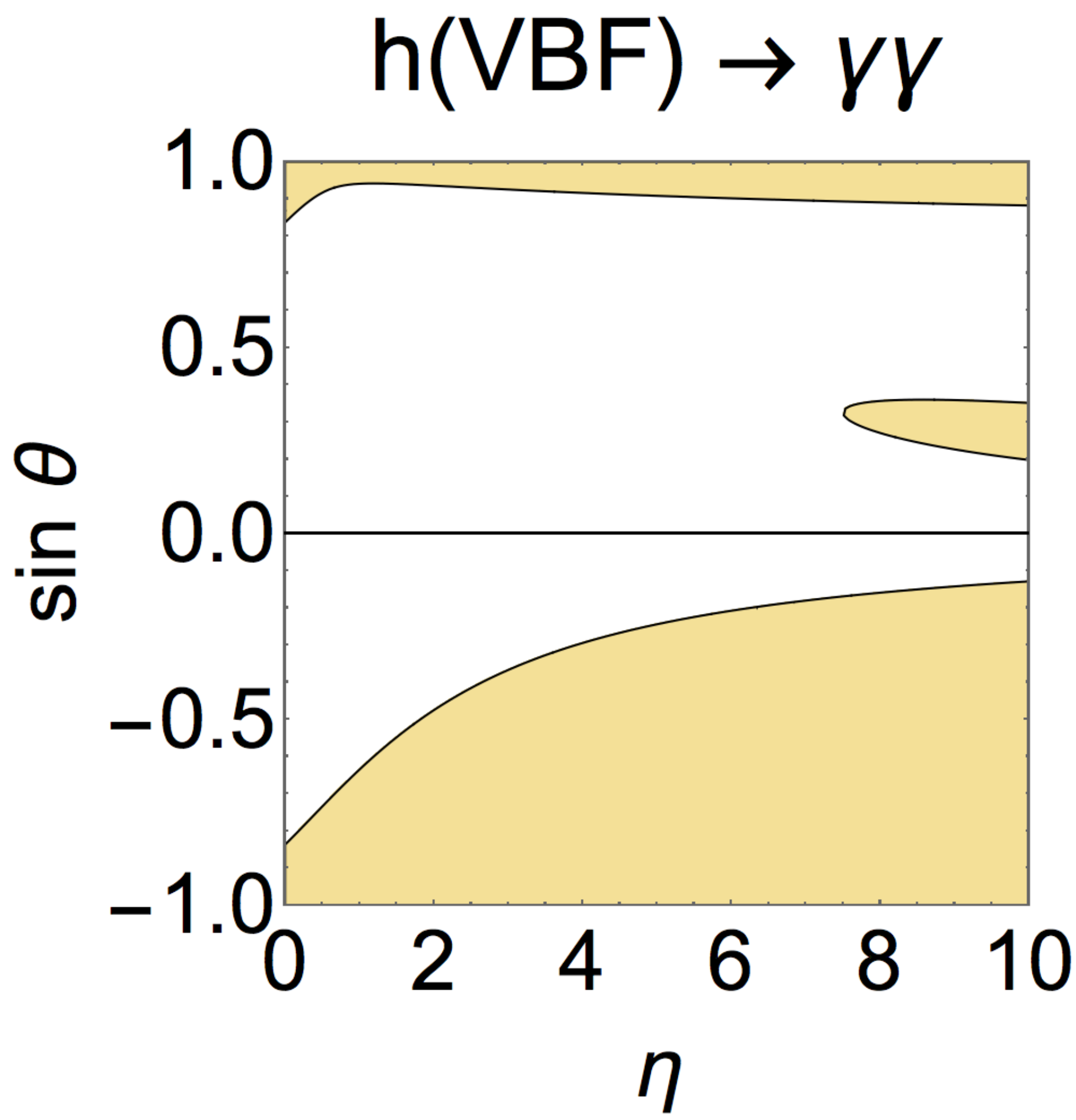} 
\includegraphics[width=0.24\textwidth]{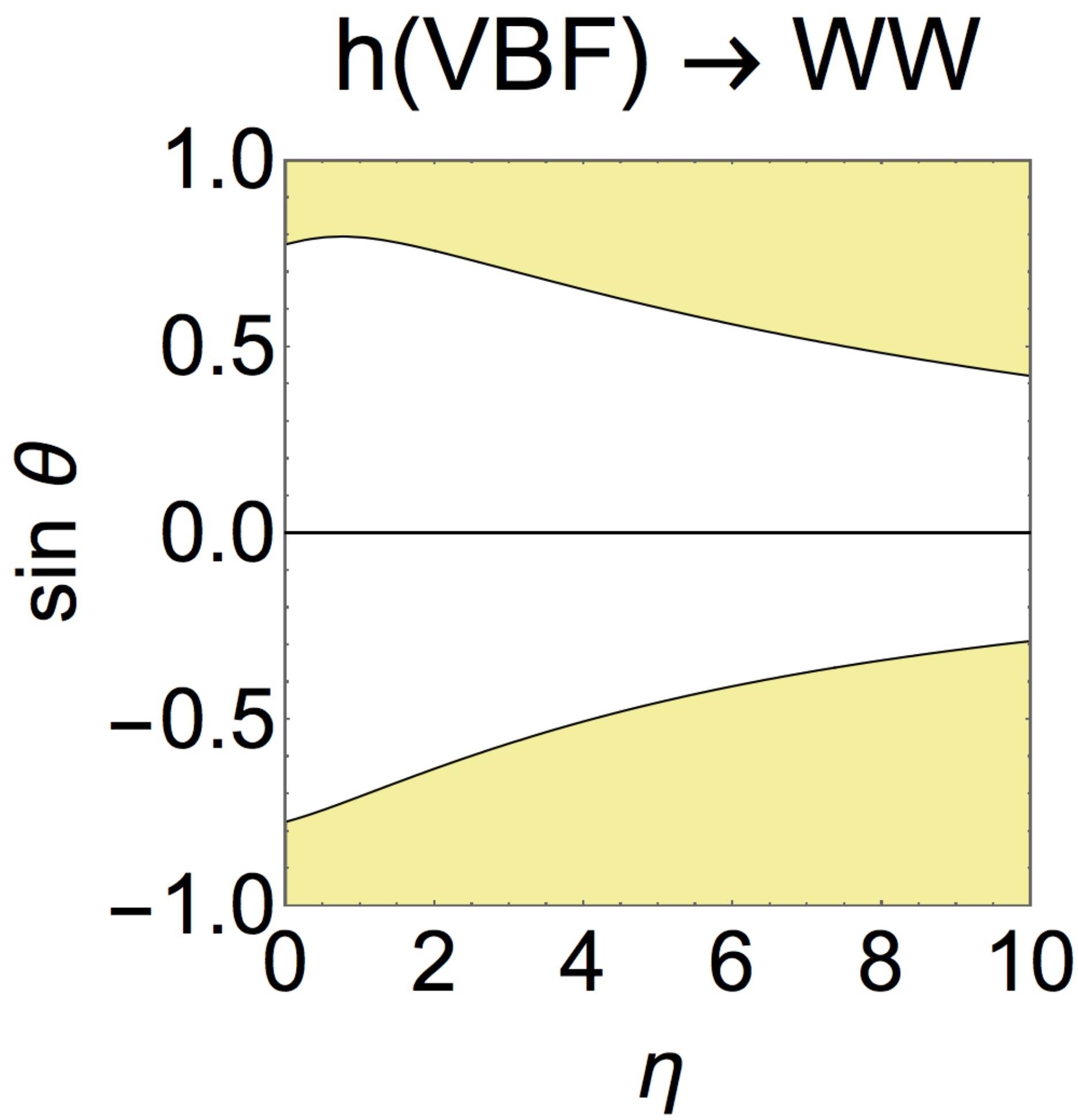} 
\includegraphics[width=0.24\textwidth]{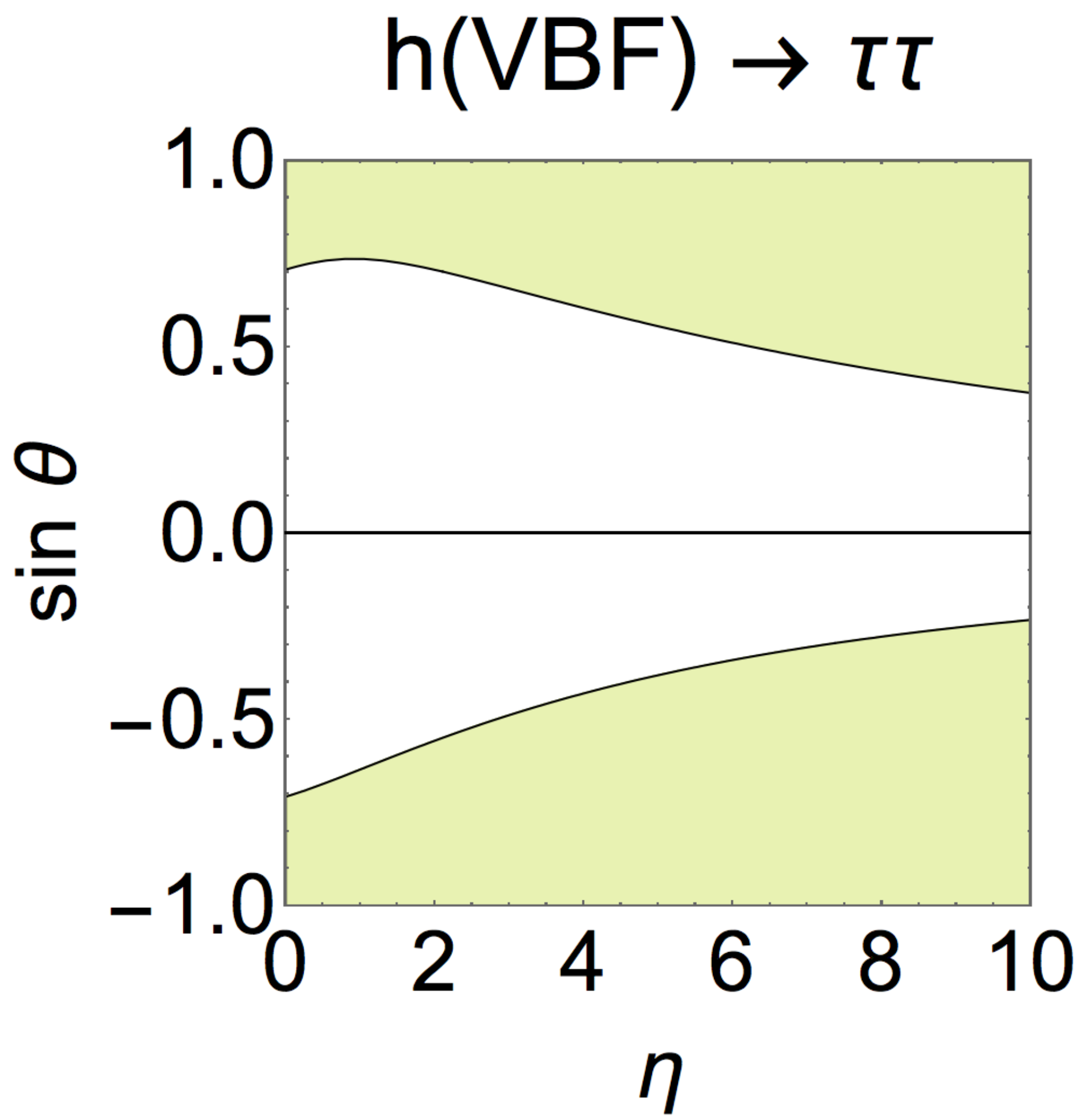}
\vspace{-5mm}
  \end{center}
\caption{
The 2$\sigma$-excluded regions from the signal strength of 125\,GeV Higgs.
The top-partner parameters are chosen as an illustration to present the contribution from each channel. 
}
	\label{each}
\end{figure}

We first examine the bound on $\theta$ and $\eta$ from the Higgs signal strengths in various channels.
The ``partial signal strength'' for the Higgs production becomes
\al{
\mu_\tx{ggF}
	&=	\paren{\cos\theta-{\Delta b_g\over b_g^\tx{top}}\eta\sin\theta}^2,&
\mu_\tx{VBF}
	=	\mu_\tx{VH}
	=	\mu_\tx{ttH}
	&=	\cos^2\theta,
}
where ggF, VBF, VH, and ttH are the gluon fusion, vector-boson fusion, associated production with vector, and that with a pair of top quarks, respectively; see e.g.\ Ref.~\cite{Khachatryan:2016vau} for details.
Similarly, the partial signal strength for the Higgs decay is
\al{
\mu_{h\to\gamma\gamma}
	&=	\paren{\cos\theta-{\Delta b_\gamma\over b_\gamma^\text{SM}}\eta\sin\theta}^2
		\paren{\frac{\Gamma_h}{\Gamma_h^\tx{SM}}}^{-1},\\
\mu_{h\to gg}
	&=	\mu_\tx{ggF} \paren{\frac{\Gamma_h}{\Gamma_h^\tx{SM}}}^{-1},\\
\mu_{h\to f\bar f,WW,ZZ}
	&=	\cos^2\theta \paren{\frac{\Gamma_h}{\Gamma_h^\tx{SM}}}^{-1},
}
where the ratio of the total widths is given by
\al{
\paren{\frac{\Gamma_h}{\Gamma_h^\tx{SM}}}
	&=   \text{Br}_{h \to \text{SM others}}^{\text{SM}} \cos^2\theta
		+\text{Br}_{h \to \gamma\gamma}^{\text{SM}} \paren{\cos\theta-{\Delta b_\gamma\over b_\gamma^\text{SM}}\eta\sin\theta}^2
		+\text{Br}_{h \to gg}^{\text{SM}} \, \mu_\tx{ggF},
}
with $\text{Br}_{h \to \text{SM others}}^{\text{SM}}=0.913$, $\text{Br}_{h \to \gamma\gamma}^{\text{SM}}=0.002$ and $\text{Br}_{h \to gg}^{\text{SM}}=0.085$.
We compare these values with the corresponding constraints given in Ref.~\cite{Khachatryan:2016vau}.
Results are shown in Fig.~\ref{combined} for the matter contents summarized in Table~\ref{table of fields}.
We note that the region near $\theta \simeq 0$ is always allowed by the signal strength constraints, though it is excluded by the di-photon search as we will see.

\subsection{Bound from $s\to ZZ\to4l$}\label{ZZ4l constraint}

\begin{figure}
\begin{center}
\includegraphics[width=0.5\textwidth]{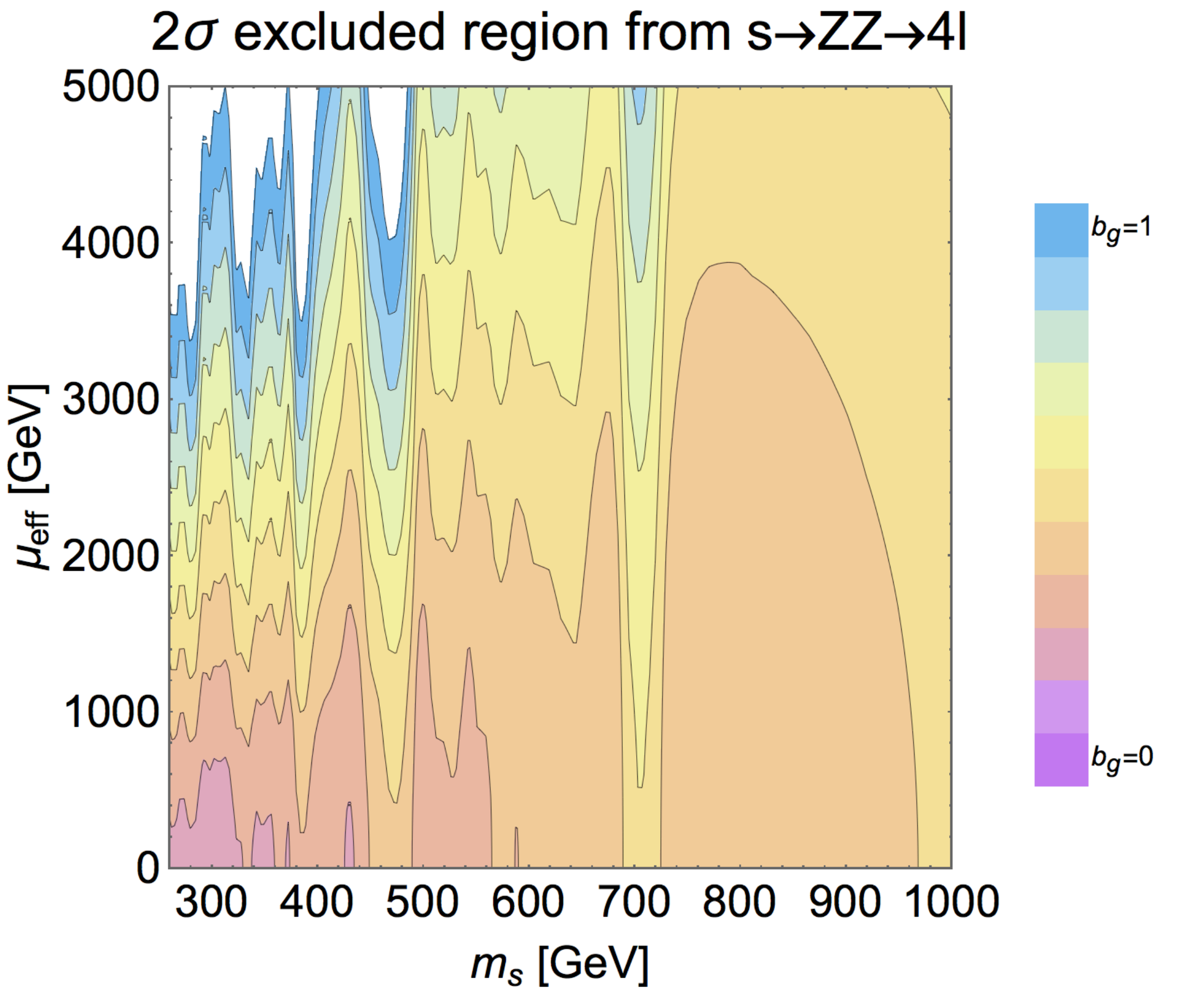}
\vspace{-5mm}
  \end{center}
\caption{
  The 2$\sigma$-excluded regions from $s\to ZZ\to4l$ bound in the $\mu_\tx{eff}$ vs $m_s$ plane. The color is changed in increments of 0.1. The weakest bound starts existing from $b_g=0.2$.
$K$-factor is set to be $K=1.6$.
}
\label{ZZ4lconstraint}
\end{figure}

One of the strongest constraints on the model comes from the heavy Higgs search in the four lepton final state at $\sqrt{s}=13\TeV$ at ATLAS~\cite{ATLAS-CONF-2016-079}.
Experimentally, an upper bound is put on the cross section $\sigma\fn{pp \to s \to ZZ\to 4l}$, with $l=e,\mu$, for each $m_s$.
Its theoretical cross section is obtained by multiplying the production cross section~\eqref{singlet production} by the branching ratio $\BR\fn{s\to ZZ}=\Gamma\fn{s\to ZZ}/\Gamma\fn{s\to\tx{all}}$ and $\paren{\BR_\tx{SM}\fn{Z\to ee,\mu\mu}}^2\simeq\paren{6.73\%}^2$; see Sec.~\ref{tree decay section}.

In Fig.~\ref{ZZ4lconstraint}, we plot $2\sigma$ excluded regions on the $\mu_\tx{eff}$ vs $m_s$ plane with varying $b_g$ from 0 to 1 with incrementation 0.2. The weakest bound starts to exist on the plane from $b_g=0.2$.
$K$-factor is set to be $K=1.6$.
The experimental bound 
becomes milder for large $\mu_\text{eff}$ because the di-Higgs channel dominate the decay of the neutral scalar.
The large fluctuation of the bound is due to the statistical fluctuation of the original experimental constraint.

We note that though we have focused on the strongest constraint at the low $m_s$ region, the other decay channels of $WW\to l\nu qq$ and of $ZZ\to\nu\nu qq$ and $ll\nu\nu$ may also become significant at the high mass region $m_s\gtrsim 700\GeV$.

\subsection{Bound from $s\to\gamma\gamma$}\label{general diphoton bound}

\begin{figure}[tp]
\begin{center}
\includegraphics[width=0.45\textwidth]{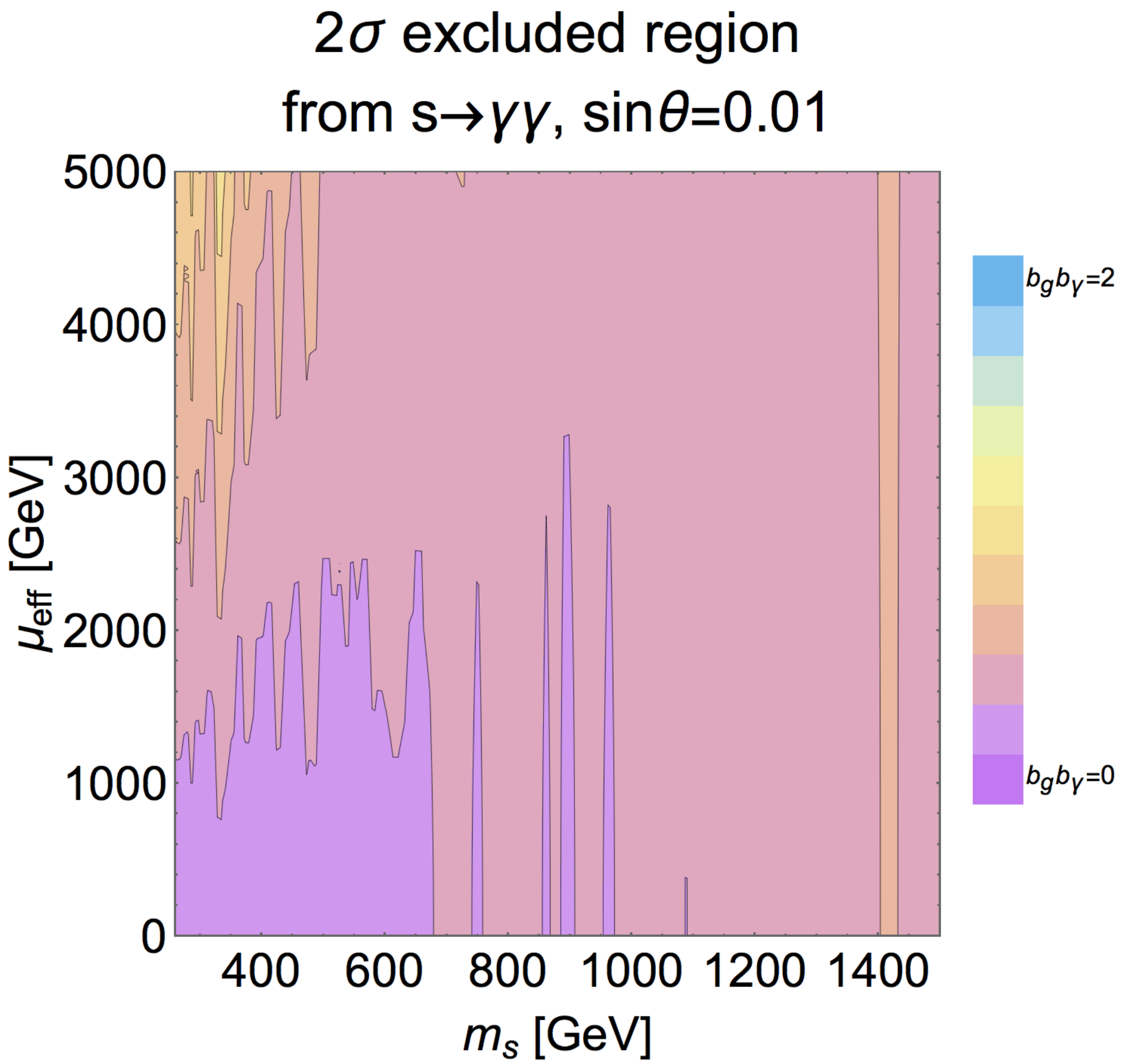}
\includegraphics[width=0.45\textwidth]{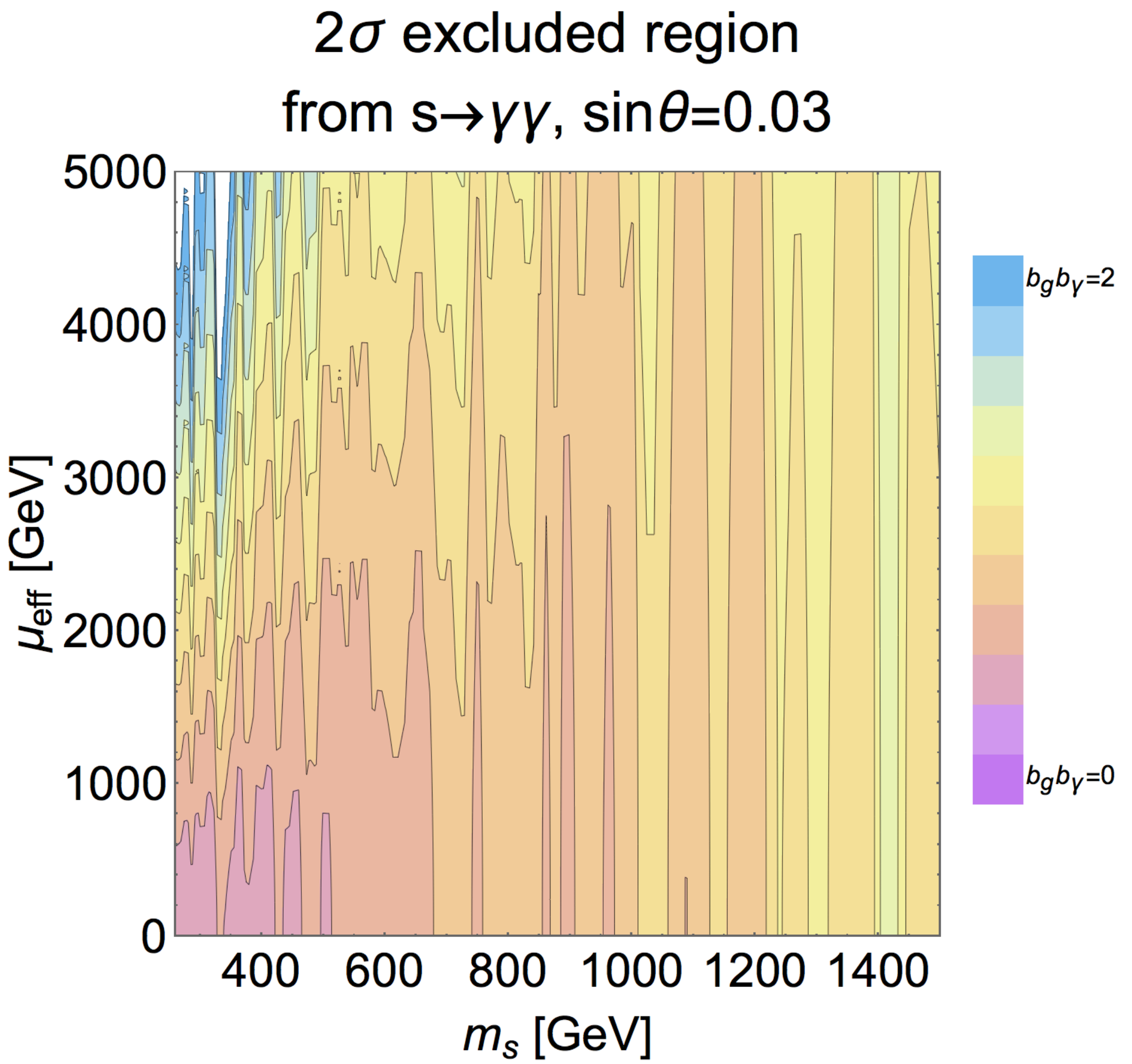}\bigskip\\
\includegraphics[width=0.45\textwidth]{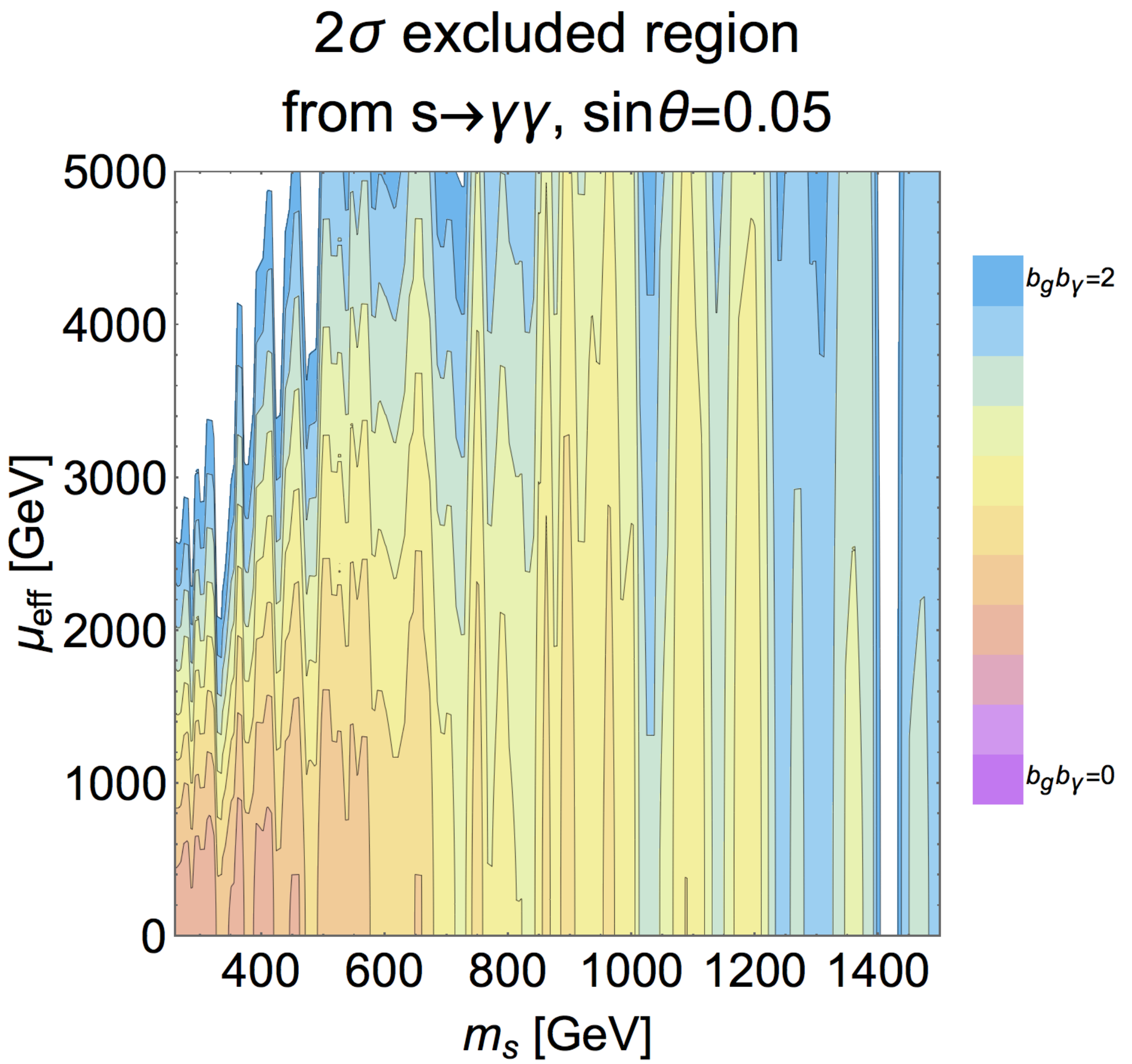}
\includegraphics[width=0.45\textwidth]{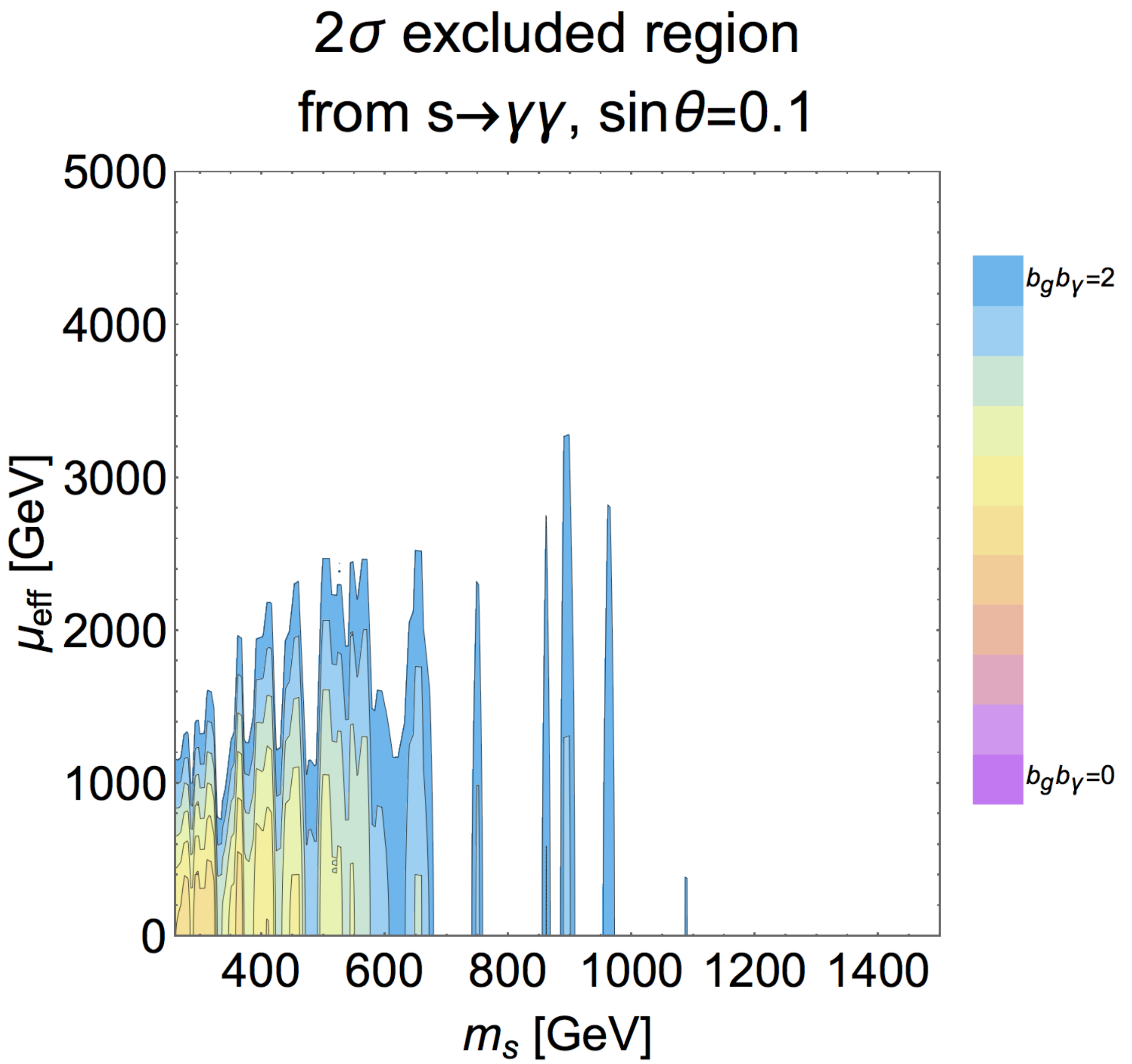}
  \end{center}
\caption{
  The 2$\sigma$-excluded regions from $s\to \gamma\gamma$ bound in the $\mu_\tx{eff}$ vs $m_s$ plane for various $\sin\theta$.
The color is changed in increments of 0.2.
$K$-factor is set to be $K=1.6$.
\label{diphoton mu vs ms}
}
\end{figure}

\begin{figure}[tp]
\begin{center}
\includegraphics[width=0.496\textwidth]{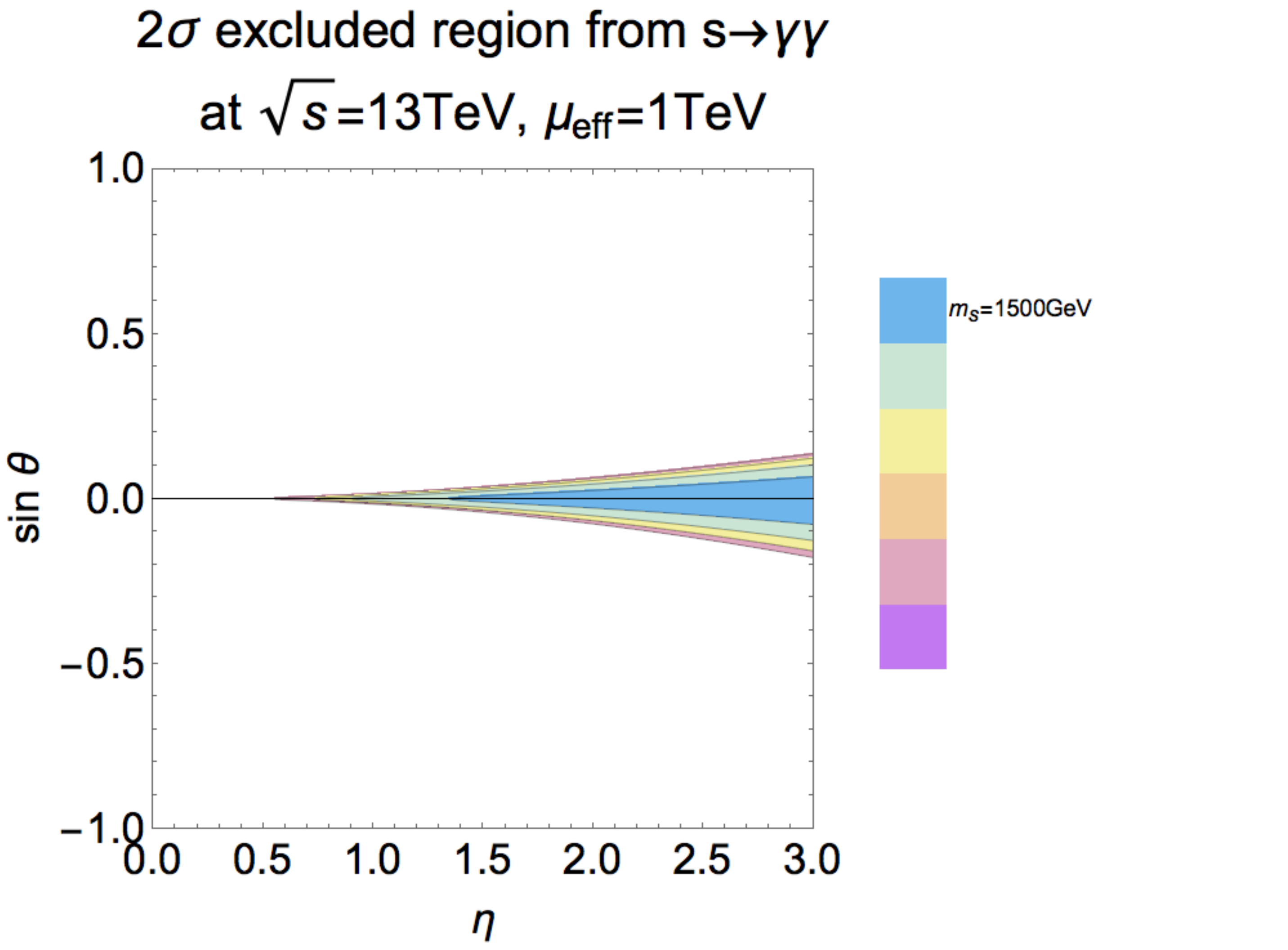}
\includegraphics[width=0.496\textwidth]{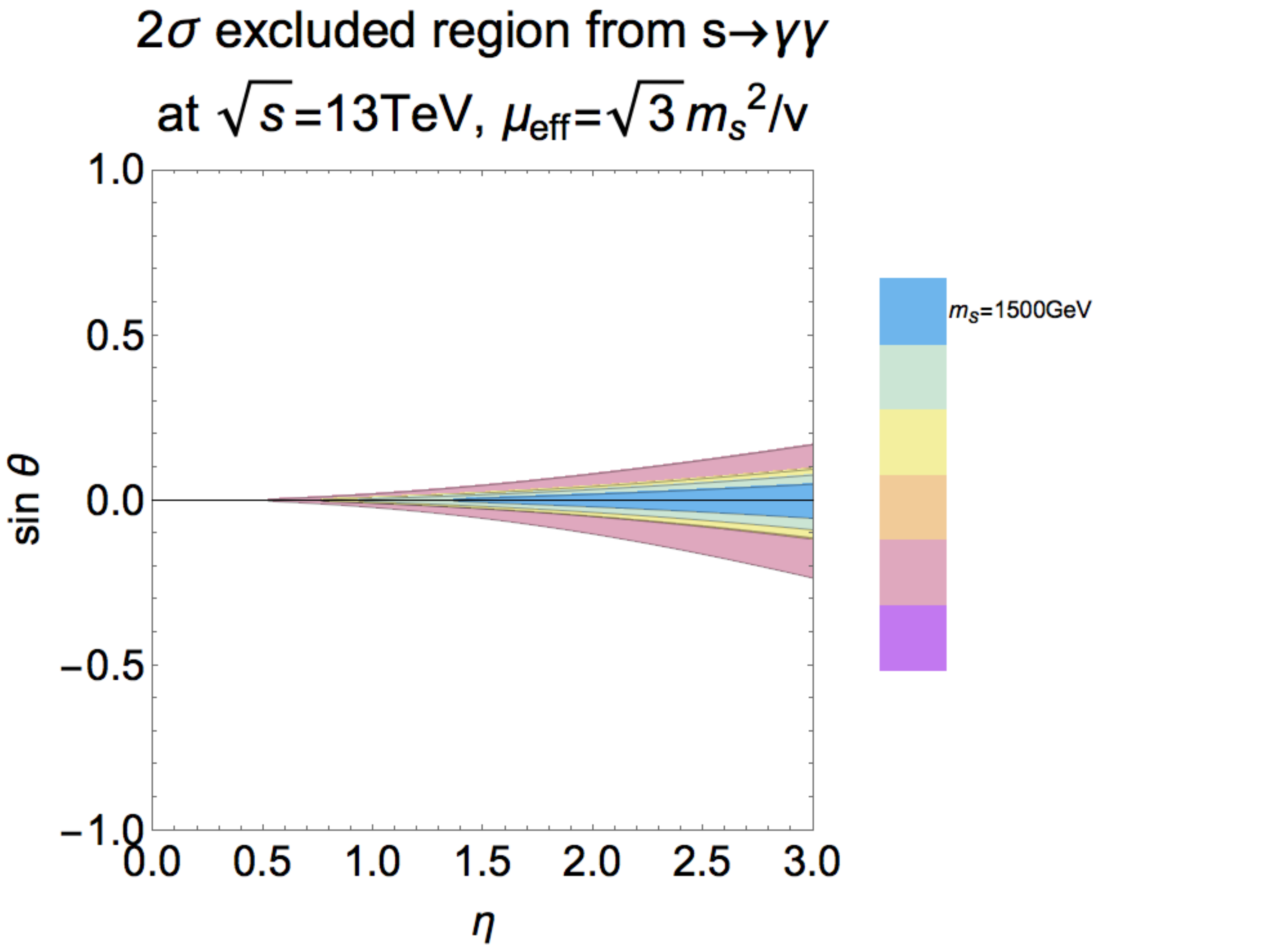}
  \end{center}
\caption{
  The 2$\sigma$-excluded regions from $s\to \gamma\gamma$ bound in the $\sin\theta$ vs $\eta$ plane for various $m_s$ with $\mu_\text{eff}=$ 1 TeV and $\sqrt{3}m_s^2/v$.
	The color is changed in increments of 300\,GeV.
	$K$-factor is set to be $K=1.6$.
	\label{diphoton sintheta vs eta}
}
\end{figure}

A strong constraint comes from the heavy Higgs search in the di-photon final state at $\sqrt{s}=13\TeV$ at ATLAS~\cite{ATLAS-CONF-2016-059}.
Experimentally, an upper bound is put on the cross section $\sigma\fn{pp \to s \to \gamma\gamma}$ for each $m_s$.
Its theoretical cross section is obtained by multiplying the production cross section~\eqref{singlet production} by the branching ratio $\BR\fn{s\to \gamma\gamma}=\Gamma\fn{s\to\gamma\gamma}/\Gamma\fn{s\to\tx{all}}$; see Sec.~\ref{tree decay section}.
Since this constraint is strong in the small mixing region, where the loop-level decay is comparable to the tree-level decay, we include the loop-level decay channels into $\Gamma\fn{s\to\tx{all}}$ for this analysis; see Sec.~\ref{loop decay}.

In Fig~\ref{diphoton mu vs ms}, we plot the $2\sigma$-excluded regions on $\mu_\tx{eff}$ vs $m_s$ plane for $\sin\theta=$0.01, 0.03, 0.05, and 0.1, with varying $b_gb_\gamma$ from 0 to 2 with incrementation 0.2.
$K$-factor is set to be $K=1.6$.
If $\sin\theta = 0.01$, broad region is excluded for $b_g b_\gamma=0.4$.
On the other hand, the experimental bound is negligibly weak in the case of $\sin\theta = 0.1$.
The large fluctuation of the bound is due to the statistical fluctuation of the original experimental constraint.

In Fig.~\ref{diphoton sintheta vs eta}, we plot the same $2\sigma$-excluded regions on the $\sin\theta$ vs $\eta$ plane for $m_s=300\GeV$, $600\GeV$, $900\GeV$, $1200\GeV$, and $1500\GeV$.
In the left and right panels, we set $\mu_\tx{eff}=1\TeV$ and $\mu_\tx{eff}=\sqrt{3}m_s^2/v$.
The latter corresponds to $\Gamma\fn{s\to hh}=\sum_{V=W,Z}\Gamma\fn{s\to VV}$ which is chosen such that there are sizable di-Higgs event and that $\mu_\tx{eff}$ is not too large.
$K$-factor is set to be $K=1.6$.
We emphasize that the small mixing limit $\sin\theta\to 0$ is always excluded by the di-photon channel in contrast to the other bounds, though it cannot be seen in Fig.~\ref{diphoton sintheta vs eta} in the small $\eta$ region due to the resolution.

The bound from $s\to Z\gamma$ is weaker and we do not present the result here.

\subsection{Bound from direct search for colored particles}
We first review the mass bound on the extra colored particles.
For the $SU(2)_L$ singlet $T$ and $B$~\cite{Aad:2015kqa,ATLAS-CONF-2016-101},
\al{
M_T,M_B\gtrsim800\GeV.
}
The mass bound for the leptoquark $\phi_{\bs 3}$, diquark $\phi_{\bs 6}$, and coloron $\phi_{\bs 8}$ are given in Refs.~\cite{Aaboud:2016qeg,Khachatryan:2016jqo}, \cite{Chivukula:2015zma}, and \cite{Sirunyan:2016iap} as
\al{
m_{\phi_{\bf 3}}
	&\gtrsim0.7\tx{--}1.1\TeV,&
m_{\phi_{\bf 6}}
	&\gtrsim 7 \TeV.&
m_{\phi_{\bf 8}}
	&\gtrsim 5.5 \TeV.
}
respectively, depending on the possible decay channels.

For the top-partner $M_T\gtrsim800\GeV$ with $\theta\simeq0$, we get $\eta\lesssim 0.3y_TN_T$. Therefore, we need rather large Yukawa coupling $y_T\simeq 2.2$ for $N_T=1$ in order to account for Eq.~\eqref{excess} by Eq.~\eqref{pp to s cross section}.\footnote{
Strictly speaking, the bound on $M_T$ slightly changes when $N_T\geq2$, and hence the bound for $y_TN_T$ could be modified accordingly.
}
The same argument applies for the bottom partner since it has the same $\Delta b_g=2/3$.

Similarly for a colored scalar with $M_\phi\gtrsim0.7$, 1.1, 5.5, and 7\,TeV, we get $\eta\lesssim \kappa_\phi N_\phi{f\ov2\TeV}$, $\kappa_\phi N_\phi{f\ov4.9\TeV}$, $\kappa_\phi N_\phi{f\ov 123 \TeV}$, and $\kappa_\phi N_\phi{f\ov 200 \TeV}$, respectively.
For $\theta \simeq 0$, the value of $b_g$ is suppressed or enhanced by extra factors ${1\over6}/{2\over3}=1/4$, ${5\ov6}/{2\ov3}={5\ov4}$, and $1/{2\over3}=3/2$, respectively, compared to the top partner. Therefore, from Eq.~\eqref{eta required}, we need $\kappa_\phi N_\phi f\gtrsim5$--13\,TeV, {106}\,TeV, and {54}\,TeV for $\phi_{\bs 3}$, $\phi_{\bs 6}$, and $\phi_{\bs 8}$, respectively, in order to account for the 2.4$\sigma$ excess at $\theta^2\ll1$.

\section{Accounting for 2.4$\sigma$ excess of $b\bar b\gamma\gamma$ by $m_s=300\GeV$}\label{2.4sigma}

\begin{figure}[tp]
\begin{center}
\includegraphics[width=0.4\textwidth]{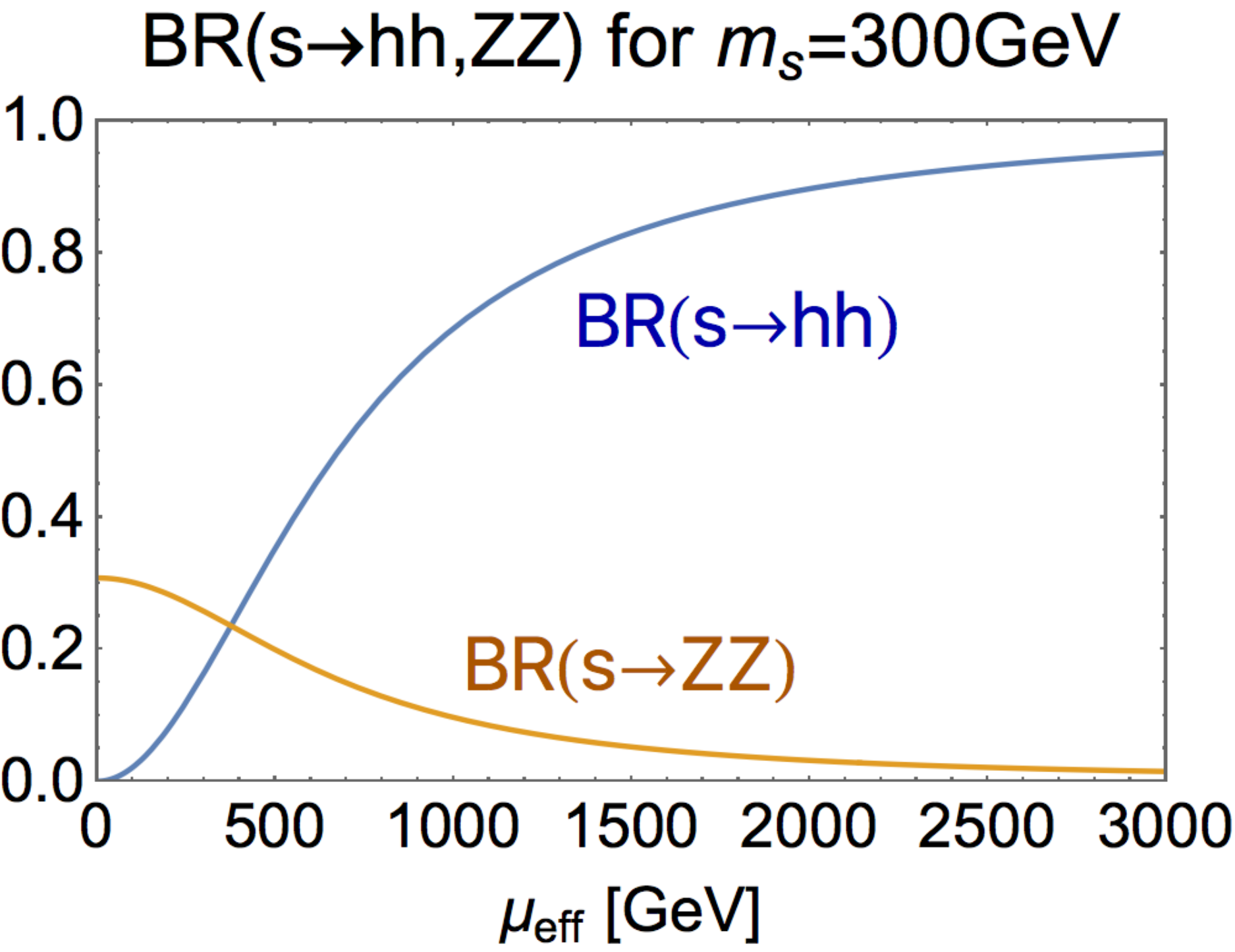}
  \end{center}
\caption{Branching ratios $\BR\fn{s\to hh}$ and $\BR\fn{s\to ZZ}$ at $m_s=300\GeV$ as functions of $\mu_\tx{eff}$.
}\label{BR300}
  \end{figure}

It has been reported by the ATLAS Collaboration that there exist 2.4$\sigma$ excess of $hh$-like events in the $b\bar b\gamma\gamma$ final state~\cite{Aad:2014yja}.
This corresponds to the extra contribution to the SM cross section\footnote{
At $\sqrt{s}=8\TeV$, $\sigma_\tx{SM}\fn{pp\to hh}=9.2\fb$.
The expected number of events are $1.3\pm0.5$, $0.17\pm0.04$, and $0.04$ for the non-$h$ background, single $h$, and the SM $hh$ events, respectively.
Since the observed number of events is 5, excess is $5-1.3-0.17=3.5$, which is $3.5/0.04=87.5$ times larger than the SM $hh$ events.
Therefore, the excess corresponds to $9.2\fb\times87.5=0.8\pb$.
}
\al{
\sigma\fn{pp\to hh}_\tx{extra,\,8\TeV}
	&\simeq 0.8\pb.
		\label{excess}
}
In Fig.~\ref{BR300}, we plot the branching ratio at $m_h=300\GeV$ as a function of $\mu_\tx{eff}$.

\subsection{Signal}

With $m_ s=300\GeV$, we get the luminosity functions
\al{
\left.\tau{\df\mc L^{gg}\over\df\tau}\right|_{m_s=300\GeV}
	&\simeq	\begin{cases}
				17.2	&	(\sqrt{s}=8\TeV),\\
				54.5	~(64.2)&	(\sqrt{s}=13~ (14)\TeV),\\
				263 ~(357)&  (\sqrt{s}=28~ (33) \TeV),\\
				2310~(1470)	&	(\sqrt{s}=100~(75)\TeV).
			\end{cases}
			\label{luminosity function}
}
%
%
That is,
\al{
\sigma\fn{pp\to s}_{m_s=300\GeV}
 	&\simeq
		\sqbr{b_g\over-1/3}^2\sqbr{\alpha_s\over0.1}^2\sqbr{K\over1.6}\times
		\begin{cases}
				1.0\pb	&	(\sqrt{s}=8\TeV),\\
				3.2~(3.8)\pb	&	(\sqrt{s}=13~(14)\TeV),\\
				15~(18)\pb        &      (\sqrt{s}=28~(33)\TeV),\\
				130~(83)\pb	&	(\sqrt{s}=100~(75)\TeV).
		\end{cases}
 		\label{pp to s cross section}
}

\begin{figure}
\begin{center}
\includegraphics[width=0.24\textwidth]{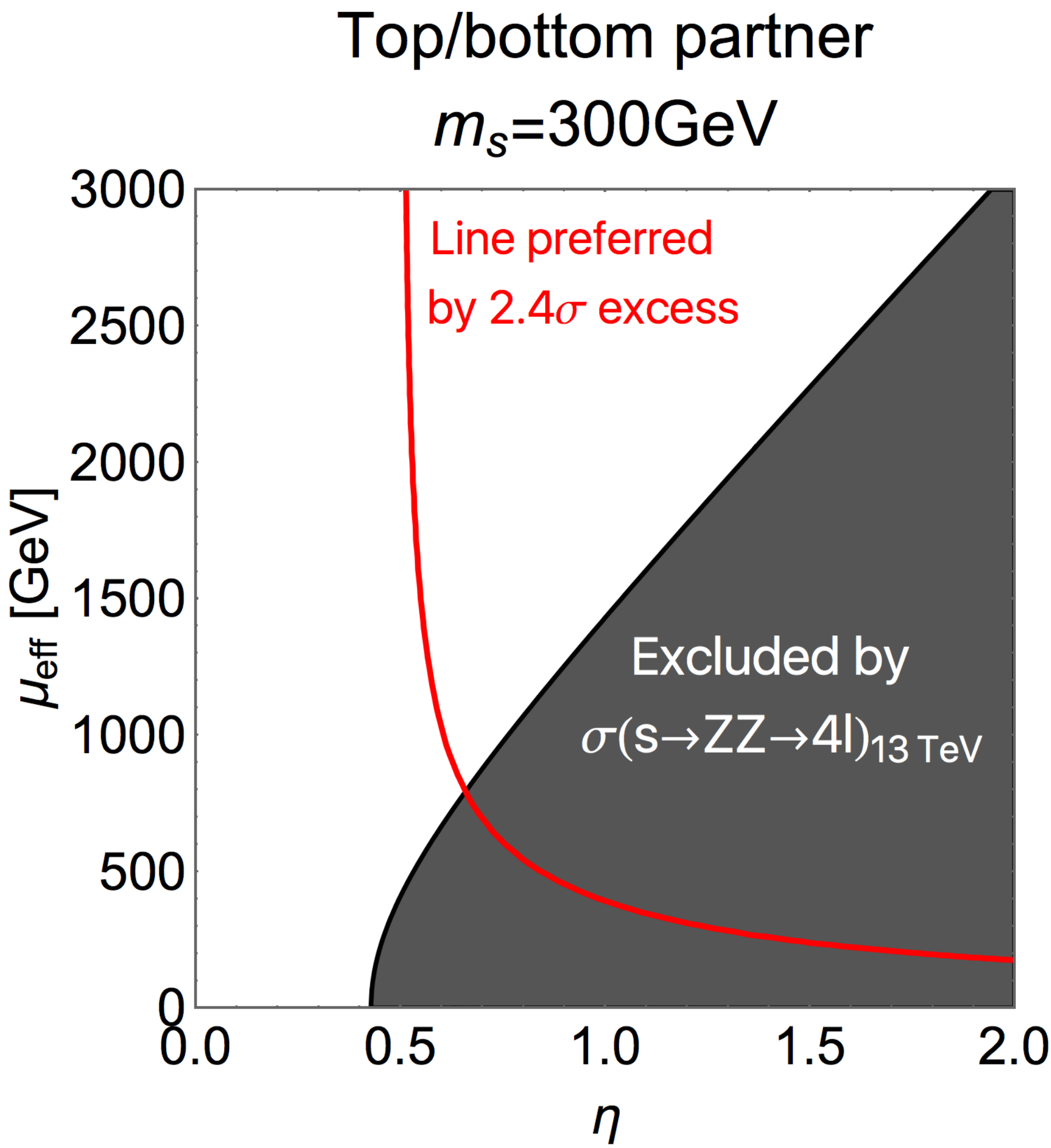}
\includegraphics[width=0.24\textwidth]{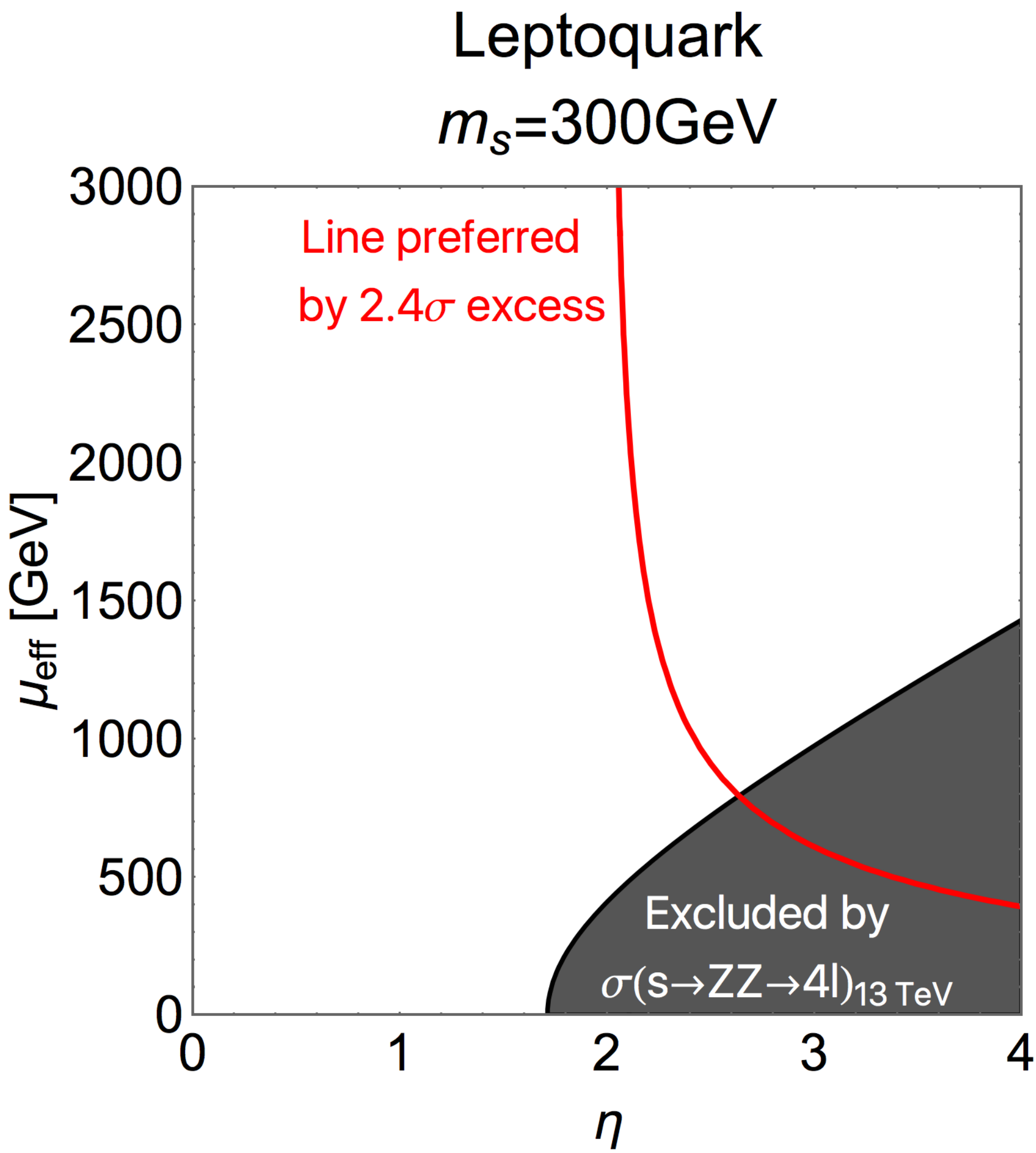}
\includegraphics[width=0.24\textwidth]{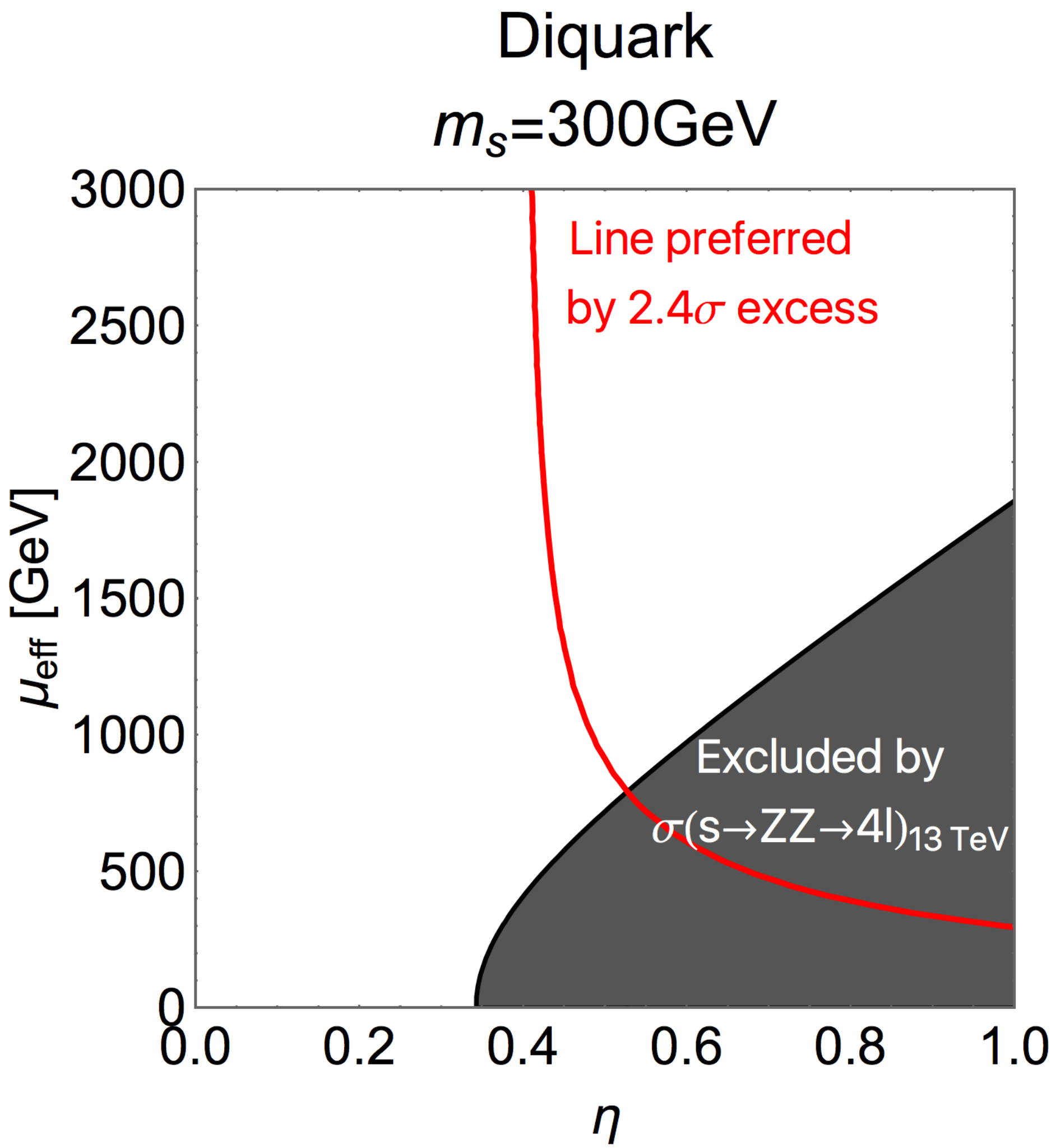}
\includegraphics[width=0.24\textwidth]{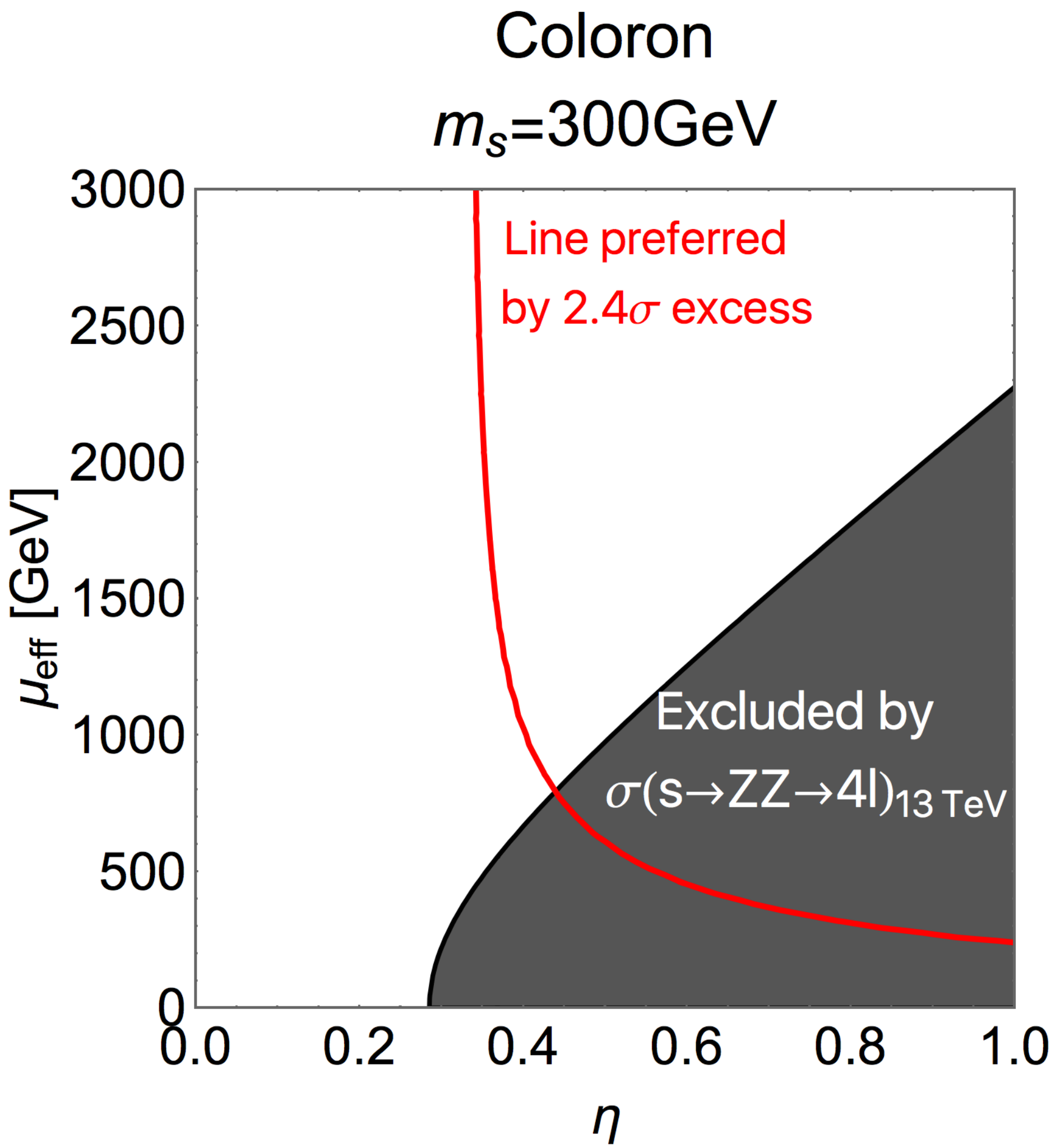}
  \end{center}
\caption{In each panel, the line corresponds to the preferred contour to explain the 2.4$\sigma$ excess at $m_s=300\GeV$, and the shaded region is excluded at the 95\% C.L.\ by $\sigma\fn{ZZ\to4l}_{13\TeV}$.
The $K$-factor is set to be $K=1.6$.
The region $10^{-4}\lesssim\theta^2\ll1$ is assumed.
Note that the plotted region of $\eta$ in horizontal axis differs panel by panel.
}\label{excess fig}
  \end{figure}

In Fig.~\ref{excess fig}, we plot the preferred contour to explain the 2.4$\sigma$ excess at $m_s=300\GeV$, where the shaded region is excluded at the 95\% C.L.\ by the $\sigma\fn{pp \to s \to ZZ\to4l}_{13\TeV}$ constraint that has been discussed in Sec.~\ref{ZZ4l constraint}. We have assumed the $K$-factor $K=1.6$.

We see that at the benchmark point $\theta\simeq0$, the lowest and highest possible values of $\mu_\tx{eff}$ and $\eta$ are, respectively,
\al{
\mu_\tx{eff}
	&\gtrsim 800\GeV,	&
\eta
	&\lesssim \begin{cases}
				0.66	&	\tx{top/bottom partner,}\\
				2.6		&	\tx{leptoquark,}\\
				0.53	&	\tx{diquark,}\\
				0.44	&	\tx{coloron,}
				\end{cases}	
				\label{eta required}
}
in order to account for the cross section~\eqref{excess}.
The ratio of the upper bound on $\eta$ is given by the scaling $\propto\paren{\Delta b_g}^2$.

\subsection{Constraints}
When $m_s = 300\GeV$, the $95\%$ C.L.\ upper bound at $\sqrt{s} = 13\TeV$ is $\sigma\fn{s(\tx{ggF})\to ZZ\to 4l}_{13\TeV}\lesssim 0.8\fb$~\cite{ATLAS-CONF-2016-079}; see also Fig.~\ref{ZZ4lconstraint}. The corresponding excluded region is plotted in Fig.~\ref{excess fig}.

Currently, the strongest direct constraint on the di-Higgs resonance at $m_s=300\GeV$ comes from the $\sqrt{s}=8\TeV$ data in the $b\bar b\gamma\gamma$ final state at CMS~\cite{diHiggsCMS} and in $b\bar b\tau\tau$ at ATLAS~\cite{diHiggsATLAS}:
\al{
{\sigma\fn{pp\to s\to hh}}_\tx{8\TeV}
	<	\begin{cases}
			1.1\pb	&	\tx{($b\bar b\gamma\gamma$ at CMS)},\\
			1.7\pb	&	\tx{($b\bar b\tau\tau$ at ATLAS)},
		\end{cases}
			\label{8TeV bound}
}
at the 95\% C.L.
The preferred value~\eqref{excess} is still within this limit.

We note that the current limit for the $m_s=300\GeV$ resonance search at $\sqrt{s}=13\TeV$ is from the $b\bar b\gamma\gamma$ final state at ATLAS~\cite{diHiggsATLAS} and from $b\bar bbb$ at CMS~\cite{diHiggsCMS}:
\al{
{\sigma\fn{pp\to s\to hh}}_\tx{13\TeV}
	<	\begin{cases}
			5.5\pb	&	\tx{($b\bar b\gamma\gamma$ at ATLAS)},\\
			11\pb	&	\tx{($b\bar bbb$ at CMS)},
		\end{cases}
}
at the 95\% C.L.
This translates to the $\sqrt{s}=8\TeV$ cross section:
\al{
{\sigma\fn{pp\to s\to hh}}_\tx{8\TeV}
	<	\begin{cases}
			1.7\pb	&	\tx{($b\bar b\gamma\gamma$ at ATLAS)},\\
			3.5\pb	&	\tx{($b\bar bbb$ at CMS)}.
		\end{cases}
}
This is weaker than the direct 8\,TeV bound~\eqref{8TeV bound}.

The branching ratio for $s\to\gamma\gamma$ is\footnote{
The power of $m_s$ dependence is valid in the limit $m_s\gg 2m_h$.
}
\al{
\BR\fn{s\to\gamma\gamma}
	&\sim
		2.3\times10^{-3}\sqbr{\alpha\over1/129}^2\sqbr{\mu_\tx{eff}\over800\GeV}^{-2}\sqbr{b_\gamma\over-8/9}^2\sqbr{m_s\over300\GeV}^4\sqbr{\sin\theta\over0.01}^{-2}.
}
We see that the loop suppressed decay into diphoton is negligible compared to the tree-level decay via the interaction~\eqref{shh coupling}.
For $m_s=300$\,GeV, the cross section at $\sqrt{s}=13\,\tx{TeV}$ is
\al{
 \sigma\fn{pp\to s \to \gamma\gamma}_\tx{13\TeV}
 	\sim7.4\fb
 \sqbr{b_g\over-1/3}^2\sqbr{b_\gamma\over-8/9}^2
 \sqbr{\alpha_s\over0.1}^2\sqbr{\alpha\over1/129}^2\sqbr{\mu_\tx{eff}\over800\GeV}^{-2}\sqbr{\sin\theta\over0.01}^{-2}.
}
We see that the loop-suppressed $\Gamma\fn{s\to\gamma\gamma}$ becomes the same order as $\Gamma\fn{s\to hh}$ when $\theta\lesssim 10^{-3}$ and that the region $\theta\lesssim 10^{-2}$ is excluded by the diphoton search, $\sigma\fn{pp\to s \to \gamma\gamma}_\tx{13\TeV}\lesssim10\fb$~\cite{ATLAS-CONF-2016-059}, for a typical set of parameters that explains the 300\,GeV excess; see also Sec.~\ref{general diphoton bound}.

We comment on the case where the neutral scalar is charged under the $Z_2$ symmetry, $S\to-S$, or is extended to a complex scalar charged under an extra U(1), $S\to e^{i\varphi}S$.
In such a model,
the effective coupling in the small mixing limit becomes
\begin{align}
 \mu_\text{eff} &\sim \frac{m_sf}{v}\lesssim {m_s\ov\eta};
	\label{Z2 relation}
\end{align}
see Appendix~\ref{Z2 model section}.
That is, for a given $m_s$, there is an upper bound on the product $\mu_\tx{eff}\,\eta$: $\mu_\tx{eff}\,\eta\lesssim m_s$.
On the other hand, the production cross section and the di-Higgs decay rate of $s$ are proportional to $\eta^2$ and $\mu_\tx{eff}^2$, and hence there is a preferred value of $\mu_\tx{eff}\,\eta$ in order to account for the 2.4\,$\sigma$ excess by $m_s=300\GeV$; see Fig.~\ref{excess fig}. In the $Z_2$ model and the $U(1)$ model, this preferred value exceeds the above upper bound.
That is, they cannot account for the excess.
More rigorous proof can be found in Appendix~\ref{Z2 model section}.

On the other hand, a singlet scalar that does not respect additional symmetry does not obey this relation~\eqref{Z2 relation}.
Because of this reason, a singlet scalar without $Z_2$ symmetry is advantageous to enhance the di-Higgs signal in general, and can explain the excess by $m_s=300\GeV$.

\section{Summary and discussion}\label{summary section}
We have studied a class of models in which the di-Higgs production is enhanced by the $s$-channel resonance of the neutral scalar that couples to a pair of gluons by the loop of heavy colored fermion or scalar.
As such a colored particle, we have considered two types of possibilities: 
\begin{itemize}
\item the vector-like fermionic partner of top or bottom quark, with which the neutral scalar may be identified as the dilaton in the quasi-conformal sector,
\item the colored scalar which is either triplet (leptoquark), sextet (diquark), or octet (coloron).
\end{itemize}

We have presented the future prospect for the enhanced di-Higgs production in the LHC and beyond. Typically, the top/bottom partner models give a cross section $\sigma\fn{pp\to s}\gtrsim1\fb$, which could be accessed by a luminosity of $\mc O\fn{\tx{ab}^{-1}}$, for the scalar mass $m_s\lesssim1.3\TeV$, 2\,TeV, and 4\,TeV at the LHC, HE-LHC, and FCC, respectively.

We have examined the constraints from the direct searches for the di-Higgs signal and for a heavy colored particle, as well as the Higgs signal strengths in various production and decay channels.
Typically small and large mixing regions are excluded by the diphoton resonance search and by the Higgs signal strength bounds, respectively.
Region of small $\mu_\tx{eff}$ is excluded by the diphoton search as well as by the $s\to ZZ\to 4l$ channel.

We also show a possible explanation of the 2.4$\sigma$ excess of the di-Higgs signal in the $b\bar b\gamma\gamma$ final state, reported by the ATLAS experiment.
We have shown that the $Z_2$ model explained in Appendix~\ref{Z2 model section} cannot account for the excess, while the general model in Appendix~\ref{general model} can. 
A typical benchmark point which evades all the bounds and can explain the excess is
\al{
\mu_\tx{eff}
	&\sim 1\TeV,	&
\eta
	&\sim \begin{cases}
				0.6	&	\tx{top/bottom partner,}\\
				2.4	&	\tx{leptoquark,}\\
				0.5	&	\tx{diquark,}\\
				0.4	&	\tx{coloron,}
				\end{cases}	&
\sin\theta
	&\sim	0.1.
}
For the top/bottom partner $T,B$, the required value to explain the 2.4$\sigma$ excess for the Yukawa coupling is rather large $y_FN_F\gtrsim 2.2$, where $N_F$ is the number of $F=T,B$ introduced. For the colored scalar $\phi$, required value of the neutral scalar VEV, $f=\left\langle S\right\rangle$, are
\al{
f\kappa_\phi N_\phi\gtrsim \begin{cases}
		5\tx{--}13\TeV	& \tx{leptoquark, depending on possible decay channels},\\
		{106} \TeV      & \tx{diquark},\\
		{54} \TeV 		& \tx{coloron},
		\end{cases}
}
where $\kappa_\phi$ and $N_\phi$ are the quartic coupling between the colored and neutral scalars and the number of colored scalar introduced, respectively.

In this paper, we have restricted ourselves to the case where the colored particle running in the blog in Fig.~\ref{Di-Higgs figure} are $SU(2)_L$ singlet. Cases for doublet, triplet, etc., which could be richer in phenomenology, will be presented elsewhere.
We have assumed $M_F, M_\phi\gtrsim m_s$ to justify integrating out the colored particle. It would be worth including loop functions to extend the region of study toward $M_F, M_\phi\lesssim m_s$.
A full collider simulation of this model for HL-LHC and FCC would be worth studying. A theoretical background of this type of the neutral scalar assisted by the colored fermion/scalar is worth pushing, such as the dilaton model and the leptoquark model with spontaneous $B-L$ symmetry breaking.

\subsection*{Acknowledgement}
The work of K. Nakamura\ and K.O.\ are partially supported by the JSPS KAKENHI Grant No.~26800156 and Nos.~15K05053, 23104009, respectively.
S.C.P. and Y.Y. are supported by the National Research Foundation of Korea (NRF) grant funded by the Korean government (MSIP) (No. 2016R1A2B2016112).
K. Nishiwaki would like to thank David London and the Group of Particle Physics of Universit\'e de Montr\'eal for the kind  hospitality at the final stage of this work.
\appendix
\section*{Appendix}

\section{General scalar potential}\label{general model}
We write down the most general renormalizable potential including the SM Higgs $H$ and the singlet $S$:
\al{
V	&=	V_S+V_H+V_{SH},
}
with\footnote{
Our $\lambda_H$ differs from the conventionally used $\lambda$ by $\lambda_H=2\lambda$, with $\lambda=m_h^2/2v^2\simeq0.13$ in the SM; see e.g.\ Refs.~\cite{Hamada:2012bp,Buttazzo:2013uya}.
}
\al{
V_S	&=	{m_S^2\over2}S^2+{\mu_S\over3!}S^3+{\lambda_S\over4!}S^4,\\
V_H
	&=	m_H^2\ab{H}^2+\frac{\lambda_H}{2}\ab{H}^4,\\
V_{SH}
	&=	\mu\,S\ab{H}^2+{\kappa\over2}\,S^2\ab{H}^2,
}
where $m_S^2$ and $m_H^2$ are (potentially negative) mass-squared parameters;
$\lambda_S$, $\kappa$, $\lambda_H$ are dimensionless constants;
$\mu_S$ and $\mu$ are real parameters of the mass dimension unity;
and the tadpole term of $S$ is removed by the field redefinition $S\to S+\tx{const.}$
The $Z_2$ model corresponds to setting $\mu_S=\mu=0$ which are prohibited by the $Z_2$ symmetry: $S\to-S$.

The vacuum condition reads
\al{
\lambda_H\ab{H}^2+\mu S+{\kappa\ov2}S^2
	&=	-m_H^2,\\
\ab{H}^2\paren{\mu+\kappa S}+{\mu_S\over2}S^2+{\lambda_S\over3!}S^3
	&=	-m_S^2S.
}
Using this vacuum condition, and putting Eqs.~\eqref{H0 written} and \eqref{S written}, we can always rewrite $m_H^2$ and $m_S^2$ in terms of $v$, $f$, and other parameters.
The mixing angle can be written as
\al{
\tan2\theta
	&=	{v\paren{f\kappa+\mu}\over
			{\lambda_S\ov 3!}f^2
			-{\lambda_H\ov2}v^2
			+{\mu_S\over4}f
			-{\mu\over4}{v^2\over f}
			}.
			\label{mixing given}
}

Now the effective coupling in Eq.~\eqref{shh coupling} is written as
\al{
\mu_\tx{eff}
	&=	\paren{\kappa f+\mu}{\cos^3\theta\ov \sin\theta}
		+v\paren{3\lambda_H-2\kappa}\cos^2\theta
		+\sqbr{f\paren{\lambda_S-2\kappa}-2\mu+\mu_S}\cos\theta\sin\theta
		+\kappa v\sin^2\theta.
}
In the small mixing limit $\theta^2\ll1$, we obtain
\al{
\mu_\tx{eff}
	&=	{\kappa f+\mu\ov\theta}+v\paren{3\lambda_H-2\kappa}+\mathcal O\fn{\theta}.
}
We note that the first term also goes to a constant for fixed $v, f$ because of Eq.~\eqref{mixing given}: $\paren{\kappa f+\mu}\propto\theta$.
More explicitly,
\al{
\mu_\tx{eff}
	&\to	{\lambda_S\ov3}{f^2\ov v}+2\paren{\lambda_H-\kappa}v+{\mu_S\ov2}{f\ov v}-{\mu\ov2}{v\ov f}
		\label{explicit small mixing limit}
}
as $\theta\to0$.
That is, the $shh$ coupling vanishes in the small mixing limit:
$\mu_\tx{eff}\sin\theta\to0$.
Let us emphasize that this is a general feature since the $shh$ coupling necessarily requires the non-zero mixing term $v\,sh$ that is obtained by the replacement $h\to v$. In order to take this feature into account, we have parametrized the effective coupling as in Eq.~\eqref{shh coupling}.

The mass eigenvalues satisfy the relations,
\al{
m_s^2+m_h^2
	&=	\lambda_Hv^2+{\lambda_S\over3}f^2-{\mu\over2}{v^2\over f}+{\mu_S\over2}f,
		\label{masssum}\\
m_s^2m_h^2
	&=	\paren{fv}^2\sqbr{
			{\lambda_S\lambda_H\over3}
			-\kappa^2
			-{\mu\over f}\paren{2\kappa
				-\frac{\lambda_H}{2}{\mu_S\over\mu}
				+{\mu\over f}
				+\frac{\lambda_H}{2}{v^2\over f^2}
			}
			},	\label{massprod}
}
where we suppose $m_s>m_h\simeq125\GeV$.
The tachyon free condition is that the right hand sides of Eqs.~\eqref{masssum} and \eqref{massprod} are positive.
Also, from the condition that the quartic terms are positive in the large field limit for any linear combination of two fields, we obtain
$\lambda_S>0$, $\lambda_H>0$, and ${\paren{{\lambda_H\lambda_S\over3}-\kappa^2}\paren{{\lambda_H\ov2}+{\lambda_S\over3!}-\kappa}}>0$.\footnote{
When we allow higher dimensional operators such as $S^6$, this vacuum stability condition can be violated. In this analysis, we restrict ourselves within the potential up to quartic order terms, and assumes that this condition is met.
}

In the model without the $Z_2$ symmetry, we can remove the parameters $\mu$ and $\mu_S$ using Eqs.~\eqref{masssum} and \eqref{massprod}.
Then the mixing angle~\eqref{mixing given} may be rewritten as
\al{
\tan2\theta
	&=	{\sqrt{\lambda_Hv^2-m_h^2}\sqrt{m_s^2-\lambda_Hv^2}\ov{m_s^2+m_h^2\ov2}-\lambda_Hv^2}.
		\label{tan2th by masses}
}
Such a solution for $\lambda_H>0$ exists when only when
\al{
{m_h^2\ov v^2}<\lambda_H<{m_s^2\ov v^2}.
	\label{condition on lambdaH}
}
We see that the small mixing limit corresponds to $\lambda_H\searrow m_h^2/v^2$.
Also one may remove $\mu,\mu_S$ from the small mixing limit~\eqref{explicit small mixing limit}:
\al{
\mu_\tx{eff}
	&\to	v\paren{
				\paren{\lambda_H-2\kappa}
				+{m_s^2+m_h^2\over v^2}}
	=		v\paren{
				-2\kappa
				+{m_s^2+2m_h^2\over v^2}},
				\label{small mixing limit in general}
}
where we used Eqs.~\eqref{masssum} and \eqref{massprod} in the first step, and substitued the $\lambda_H\searrow m_h^2/v^2$ limit in the next step.
We see that the Higgs-singlet mixing $\kappa$ remains to be a free parameters even in the small mixing limit.

If we want to explain the $b\bar b\gamma\gamma$ excess~\cite{Aad:2014yja}, we set $m_ s\simeq 300\,\tx{GeV}$, and get
\al{
{m_ s^2+m_h^2\over2}
	&\simeq	\paren{230\,\tx{GeV}}^2,	&
m_ s^2m_h^2
	&\simeq
		\paren{190\,\tx{GeV}}^4.
}
Even in the small mixing limit, $\mu_\tx{eff}$ in Eq.~\eqref{small mixing limit in general} with $m_s=300\GeV$ can be as large as $\mu_\tx{eff}\simeq 1\TeV$ ($2\TeV$) for $\kappa=-1$ ($-3$), which is well within the current experimental bound; see Fig.~\ref{excess fig};
if we are happy with an extremely large value, say, $\kappa=-4\pi$, we may push it up to $\mu_\tx{eff}\simeq6.7\TeV$.

\section{$Z_2$ model}\label{Z2 model section}
We consider the $Z_2$ model with $\mu=\mu_S=0$. The discussion is parallel to Appendix~\ref{general model}. The mixing angle reads
\al{
\tan2\theta
	&=	{\kappa v\over
			{\lambda_S\ov 3!}f
			-{\lambda_H\ov2}{v^2\ov f}
			}.
			\label{tan2th in Z2}
}
Especially in the limit $v\ll f$, we get $\tan2\theta\to{6\kappa\ov\lambda_S}{v\ov f}$.
Eqs.~\eqref{masssum} and \eqref{massprod} may be solved e.g.\ as
\al{
f^2	&=	{\paren{\lambda_Hv^2-m_h^2}\paren{m_s^2-\lambda_Hv^2}\ov\kappa^2v^2},&
\lambda_S
	&=	{3\kappa^2v^2\paren{m_h^2+m_s^2-\lambda_Hv^2}\ov\paren{\lambda_Hv^2-m_h^2}\paren{m_s^2-\lambda_Hv^2}}.
		\label{f and lambda in Z2}
}
For $\lambda_H>0$, the solution with $m_s>m_h>0$ again exists when and only when the condition~\eqref{condition on lambdaH} is met. This condition also ensures $\lambda_S$ to be positive.
Putting Eq.~\eqref{f and lambda in Z2} into Eq.~\eqref{tan2th in Z2}, we again obtain Eq.~\eqref{tan2th by masses}.

Finally, the small mixing limit of the effective coupling becomes
\al{
\mu_\tx{eff}
	&\to	v\paren{\lambda_H+{m_s^2+m_h^2\over v^2}}
	=		{m_s^2+2m_h^2\over v}.
		\label{Z2 mu eff small mixing}
}
If we want to set $m_s=300\GeV$, we get $\mu_\tx{eff}\simeq490\GeV$ in the small mixing limit $\theta^2\ll1$, which is already excluded by the $s\to ZZ\to4l$ search; see Fig.~\ref{excess fig}.
The $Z_2$ model cannot explain the 2.4\,$\sigma$ excess reported by ATLAS.
For larger values of $m_s$, the $Z_2$ model is still viable.

\section{Yukawa interaction between colored scalar and SM particles}\label{colored scalar section}
For the scalar in the fundamental representation $\phi_{\bs 3}$, the possible Yukawa interactions are
\al{
&\paren{\phi_{\bs 3}}^*\ol{\paren{q_\L}^c}_i\cdot l_\L^i,	&
&\paren{\phi_{\bs 3}}^*\ol{\paren{u_\R}^c}e_\R,	&
&\paren{\phi_{\bs 3}}^*\ol{\paren{d_\R}^c}e_\R,
}
depending on the hypercharge of $\phi_{\bs 3}$: $-1/3$, $-1/3$, and $-4/3$, respectively.
The superscript ${}^{c}$ denotes the charge conjugation.

We note that we can in principle write down the following diquark interactions:
\al{
&\epsilon^{abc}\epsilon^{ij}\paren{\phi_{\bs 3}}_a \ol{\paren{q_\L}^c}_{bi} \paren{q_\L}_{cj}, &
&\epsilon^{abc} \paren{\phi_{\bs 3}}_a \ol{\paren{u_\R}^c}_b \paren{u_\R}_c, &
&\epsilon^{abc} \paren{\phi_{\bs 3}}_a \ol{\paren{d_\R}^c}_b \paren{d_\R}_c, &
&\epsilon^{abc} \paren{\phi_{\bs 3}}_a \ol{\paren{u_\R}^c}_b \paren{d_\R}_c, &
}
depending on the hypercharge of $\phi_{\bs 3}$: $-1/3$, $-4/3$, $2/3$ and $-1/3$, respectively,
where $a,b,c$ and $i,j$ represent the indices of the $SU(3)_C$ and $SU(2)_L$ fundamental representations, respectively, and $\epsilon$ is the totally antisymmetric tensor.
The coexistence of the leptoquark and the diquark interactions leads to rapid proton decay.
Since the diquark interactions are strongly restricted compared with the leptoquark in direct searches in hadron colliders, we focus on the situation that only the leptoquark interactions are switched on. 
The diquark interactions can be forbidden e.g.\ by the $B-L$ symmetry.

For the symmetric scalar $\phi_{\bs 6}$, a possible Yukawa is either one of
\al{
&\ol{\paren{u_\R}^c_a}\,\paren{\phi_{\bs 6}}^{*ab}\,\paren{u_\R}_b,	&
&\ol{\paren{d_\R}^c_a}\,\paren{\phi_{\bs 6}}^{*ab}\,\paren{u_\R}_b,	&
&\ol{\paren{d_\R}^c_a}\,\paren{\phi_{\bs 6}}^{*ab}\,\paren{d_\R}_b,	&
&\epsilon^{ij}\ol{\paren{q_\L}^c_{ai}}\,\paren{\phi_{\bs 6}}^{*ab}\,\paren{q_\L}_{bj},
}
depending on the hypercharge of $\phi_{\bs 6}$: $4/3$, $1/3$, $-2/3$, and $1/3$, respectively.

For adjoint scalar, a possible lowest-dimensional Yukawa is either one of
\al{
&{1\over\Lambda}\ol{u_\R}^a\paren{\phi_{\bs 8}}_a{}^b\paren{q_\L}_{bi}\epsilon^{ij}H_j,	&
&{1\over\Lambda}\ol{u_\R}^a\paren{\phi_{\bs 8}}_a{}^b\paren{q_\L}_{bi}\paren{H^*}^i,\\
&{1\over\Lambda}\ol{d_\R}^a\paren{\phi_{\bs 8}}_a{}^b\paren{q_\L}_{bi}\epsilon^{ij}H_j,	&
&{1\over\Lambda}\ol{d_\R}^a\paren{\phi_{\bs 8}}_a{}^b\paren{q_\L}_{bi}\paren{H^*}^i,
}
depending on the hypercharge of $\phi_{\bs 8}$: $0$, $-1$, $-1$, and $0$, respectively,
where we have assigned $Y_H=+1/2$ and $\Lambda$ denotes an ultraviolet cutoff scale.

\bibliographystyle{TitleAndArxiv}
\bibliography{refs,refs_additional}

\end{document}